\documentclass[man,12pt,floatsintext,natbib]{apa7}
\usepackage[nodisplayskipstretch]{setspace}
\usepackage{booktabs,multirow}
\usepackage{graphicx}
\usepackage{enumitem}
\usepackage{amsmath,amsfonts,amsthm}
\usepackage{algorithm,algpseudocode}
\usepackage{pdfpages}
\usepackage{tikz}
\usetikzlibrary{shapes,arrows.meta,decorations.markings,positioning}

\makeatletter
\renewcommand{\subsubsection}{\@startsection{subsubsection}{3}{\z@}%
    {\b@level@three@skip}{\e@level@three@skip}%
    {\normalfont\normalsize\bfseries\slshape}}
\makeatother

\DeclareRobustCommand{\bfseries}{\fontseries{b}\selectfont}
\DeclareMathAlphabet{\mathbf}{OT1}{cmr}{b}{n}

\definecolor{red1}{RGB}{228,55,55}
\definecolor{blue1}{RGB}{0,0,0}
\long\def\yl#1{\bgroup\color{blue1}#1\egroup}

\def\t{^\top}

\def\eepsilon{\boldsymbol{\epsilon}}

\def\bbeta{\boldsymbol{\beta}}

\def\ttheta{\boldsymbol{\theta}}
\def\vvartheta{\boldsymbol{\vartheta}}
\def\mmu{\boldsymbol{\mu}}

\def\ggamma{\boldsymbol{\gamma}}

\def\llambda{\boldsymbol{\lambda}}

\def\nnu{\boldsymbol{\nu}}
\def\uupsilon{\boldsymbol{\upsilon}}

\def\SSigma{\mathbf{\Sigma}}
\def\DDelta{\mathbf{\Delta}}

\def\TTheta{\mathbf{\Theta}}

\def\II{\mathbf{I}}

\def\RR{\mathbf{R}}

\def\MM{\mathbf{M}}

\def\UU{\mathbf{U}}

\def\YY{\mathbf{Y}}
\def\ZZ{\mathbf{Z}}

\def\ee{\mathbf{e}}
\def\gg{\mathbf{g}}
\def\hh{\mathbf{h}}

\def\xx{\mathbf{x}}
\def\uu{\mathbf{u}}

\def\yy{\mathbf{y}}

\def\real{\mathcal{R}}

\def\diag{\hbox{Diag}}

\def\proj{\hbox{proj}}
\def\grad{\hbox{grad}\,}

\def\ex{\mathsf{E}}
\def\pr{\mathsf{P}}

\def\cL{\mathcal{L}}
\def\cX{\mathcal{X}}
\def\cY{\mathcal{Y}}
\def\cT{\mathcal{T}}

\def\cQ{\mathcal{Q}}
\def\sN{\mathsf{N}}

\DeclareMathOperator*{\argmax}{\mathrm{arg\,max}}

\newtheoremstyle{default}
  {0pt} 
  {0pt} 
  {\itshape} 
  {} 
  {\bfseries} 
  {.} 
  {.5em} 
  {} 
\newtheoremstyle{remark}
  {0pt} 
  {0pt} 
  {} 
  {} 
  {\itshape} 
  {.} 
  {.5em} 
  {} 
\theoremstyle{default}

\theoremstyle{remark}

\algrenewcommand\algorithmicrequire{\textbf{input:}}
\algrenewcommand\algorithmicensure{\textbf{output:}}
\algblockdefx{Do}{EndDo}{\textbf{do}}{\textbf{end do}}

\setlist[itemize, 1]{leftmargin=*, topsep=0ex, itemsep=0pt, parsep=0ex, labelindent=0pt}
\setlist[itemize, 2]{leftmargin=*, topsep=0ex, itemsep=0pt, parsep=0ex, labelindent=0pt, label=$\circ$}
\setlist[itemize, 3]{leftmargin=*, topsep=0ex, itemsep=0pt, parsep=0ex, labelindent=0pt, label=-}
\setlist[enumerate, 1]{leftmargin=*, topsep=0ex, itemsep=0pt, parsep=0ex, labelindent=0pt, label=(\arabic*)}

\begin{document}
\doublespace

\title{Calibrating Bayesian Inference}
\shorttitle{Calibrated Bayes}
\authorsnames[1,2,3]{Yang Liu, Jonathan P. Williams, Jan Hannig}
\authorsaffiliations{{University of Maryland, College Park}, {North Carolina State University}, {The University of North Carolina at Chapel Hill}}
\authornote{Correspondence should be made to Yang Liu at 3304R Benjamin Bldg, 3942 Campus Dr, University of Maryland, College Park, MD 20742. Email: yliu87@umd.edu. The first author of the paper is grateful for Youjin Sung's assistance in reviewing and editing an earlier version of the manuscript.}

\abstract{Bayesian statistics has gained popularity in psychological research due to its intuitive uncertainty quantification and convenient information-updating rules. In many applications, however, prior distributions are introduced merely as instruments to facilitate computation, rather than as representations of genuine subjective belief. Consequently, relying on standard Bayesian justifications for inferential procedures becomes conceptually ungrounded. In this paper, we recommend evaluating finite-sample performance over repeated sampling of data and parameters as an alternative justification for ``pragmatic Bayes.'' We demonstrate a key vulnerability in the usual posterior-based inference: when analysts' chosen prior distribution mismatches the true parameter-generating process, Bayesian inference can be misleading. Given that this true process is rarely known in practice, we propose a safer alternative: calibrating Bayesian credible regions to achieve frequentist validity. This latter criterion is stronger and guarantees validity of Bayesian inference regardless of the underlying parameter-generating mechanism. To solve the calibration problem in practice, we propose a novel stochastic approximation algorithm. A Monte Carlo experiment is conducted and reported, in which we observe that uncalibrated Bayesian inference can be liberal under certain parameter-generating scenarios, whereas our calibrated solution consistently maintain validity. We also illustrate the proposed calibration procedure using a real-data example involving location-scale regression.}

\maketitle
\setcounter{secnumdepth}{0}

\section{Introduction}

Recent decades have seen a rise in psychological publications using Bayesian methods \citep[e.g.,][]{Kruschke2021,VanDeSchootEtAl2017,VanDeSchootEtAl2021,volpe2025}. Bayesian inference offers intuitive uncertainty quantification using posterior probability measures. Drawing on (approximate) random samples from the posterior distribution, inference on model parameters, prediction of future data, and assessment of model fit can be conveniently performed in an analytics-free fashion \citep{GelmanEtAl2013}. Off-the-shelf software for Bayesian analysis includes not only generic Markov chain Monte Carlo (MCMC) samplers like JAGS \citep{JAGS} and Stan \citep{Stan}, but also programs designed for special modeling frameworks such as M\emph{plus} \citep{Mplus} and blavaan \citep{blavaan}.

Philosophical and statistical justification of Bayesian inference is rooted in making coherent decisions under uncertainty \citep{DeGroot1970, Savage1954}: encoding \emph{a priori} knowledge by a proper probability measure and updating knowledge by Bayes' formula after observing new data. However, applications of Bayesian procedures in psychology have been largely instrumental rather than philosophical: explicit justifications linking priors to researchers' beliefs remain scarce, whereas the reliance on default or conjugate priors is prevalent. When prior specifications fail to accurately reflect genuine prior knowledge, the theoretical basis for applying the formal Bayes' rule is compromised. To accommodate the prevalence of ``pragmatic Bayes'' in reality, methodological studies of Bayesian methods focus on performance over repeated sampling of data and/or parameters, assessing the extent to which the resulting inference remains systematically valid. This approach provides a natural and straightforward basis for evaluation that aligns directly with the scientific goal of replicability \citep[for a comprehensive discussion of replicability in psychology, see][]{NosekEtAl2022}. In addition, long-run performance is widely endorsed as a fundamental requirement for all statistical procedures: for example, \citet[][p. 295]{ReidCox2015} stated that
\begin{quote}
  \dots even if an empirical frequency-based view of probability is not used directly as a basis for inference; it is unacceptable if a procedure yielding regions of high probability in the sense of representing uncertain knowledge would, if used repeatedly, give systematically misleading conclusions.
\end{quote}
Notably, this emphasis on frequentist evaluation has long been applied to Bayesian inference, resulting in a rich literature on ``calibrated Bayes'' \citep{Dawid1982, Rubin1984, Little2006}.

Based on performance, three rationales are typically offered to defend the pragmatic use of Bayesian methods. First, certain default priors, especially ``weakly informative'' and ``objective'' priors \citep[e.g.,][]{BergerEtAl2015, BergerBernardoSun2024, DattaMukerjee2004, GelmanEtAl2008}, have been shown to exhibit desirable theoretical properties or strong empirical performance in the extant literature. Second, under suitable regularity conditions, the impact of priors diminishes as the sample size increases and the resulting posterior-based inference often resembles its frequentist counterpart in large samples \cite[e.g., the Bernstein-von Mises theorem;][Chapter 10]{vanderVaart1998}. Third, sensitivity analysis is typically recommended to ensure the robustness of statistical conclusions to different choices of priors \citep[e.g.,][]{DepaoliEtAl2020, Depaoli2022, vanErpEtAl2018}.

However, the appropriateness of the above justifications is often questionable when applied to real-world psychological studies. First, the principles of ``objectivity'' and ``non-informativeness'' in defining default priors are not grounded in a single, unified framework \citep[e.g.,][]{KassWasserman1996}. Additionally, which default prior performs best is often contingent upon the data-generating model \citep{YangBerger1998}. Therefore, finding one prior that exhibits universally strong performance is likely an elusive goal. Second, substantive researchers may have to work with small samples due to research focus or practical considerations. For instance, intersectional subpopulations defined by multiple social identities are often too narrow to amass data \citep[e.g.,][]{Cole2009}. Statistical procedures based on large-sample theory can be numerically unstable or produce misleading inference in small-sample applications \citep[e.g.,][]{VM2020}. Third, prior sensitivity analysis can be inconclusive. It is almost always possible to find a pathological prior distribution, such as one concentrated sufficiently far from the original Bayesian solution, in order to overturn the original conclusion. Bayesian computation can also be too computationally expensive to be repeated a large number of times. As such, prior sensitivity analysis is typically confined to a limited, arbitrarily chosen collection of priors, offering little diagnostic value for prior specification.

To assess the finite-sample performance of specific Bayesian procedures, a large number of Monte Carlo (MC) experiments have been conducted over the past decades, in which Bayesian tests and interval/set estimators were evaluated under various data and parameter-generating mechanisms and design factors \citep[e.g., sample sizes, number of covariates, etc.][]{FinchFrench2019, McNeish2016, McNeish2017, McNeish2017a, PreacherMacCallum2002, SmidEtAl2020}. A major limitation of MC studies is that their conclusions are model- and design-specific, not readily generalizable to scenarios beyond those explicitly tested. Consequently, the credibility of findings from psychological studies using Bayesian analysis has yet to be fully established.

The present paper has two primary objectives, pedagogical and methodological, both of which aim to enhance the performance of standard Bayesian procedures when used pragmatically. \yl{Pedagogically, we borrow results from statistical decision theory \citep{Berger1985} and inferential models \citep[IMs;][]{LiuMartin2024, MartinLiu2015} to call attention to the key notion of Bayesian validity: the principle that correct statements about parameters are rarely deemed implausible under a correct prior specification, which justifies the long-run performance of Bayesian procedures \citep[e.g.,][]{Martin2022a, Martin2022b, Martin2022c}.} In the absence of prior knowledge, where the use of Bayesian methods is purely instrumental, we demonstrate that the stronger notion of frequentist validity should be pursued because it necessarily implies Bayesian validity with any prior specification.\footnote{\yl{Throughout the paper, the term ``validity'' refers specifically to statistical validity and should not be confused with the notion of validity in evaluating measurement instruments or research designs. The precise definitions of Bayesian and frequentist validity are provided in the subsection ``Statistical Validity.''}} Methodologically, we propose a computational procedure to calibrate posterior-based inference to ensure frequentist validity. Our method leverages gradient-free stochastic approximation (SA) and manifold optimization \citep[e.g.,][]{AbsilEtAl2008, Spall1992}. We apply our method to Gaussian location-scale regression \citep{Harvey1976} in an MC experiment, which demonstrates that Bayesian inference can be invalid without proper calibration.

The rest of the article is structured as follows. We first review the theoretical foundations of statistical decision theory, IMs, and statistical validity. Specifically, we elaborate on two crucial facts: (1) Bayesian inference is not guaranteed to be valid if the specified prior disagrees with true parameter-generating prior, and (2) frequentist validity ensures Bayesian validity for any true parameter-generating prior. Next, we introduce a practical computational algorithm that calibrates Bayesian inference to achieve frequentist validity. A proof-of-concept MC experiment and an empirical illustration contrast the calibrated Bayesian inference against the standard Bayesian inference (utilizing both asymptotic theory and MCMC sampling). The paper is concluded with discussions on implications, limitations, and future avenues of research.

\section{Bayesian Inference and Statistical Validity}
\subsection{Bayesian Model}
We begin with the general definition of a Bayesian model and basic notation. Denote random data and model parameters by $\YY\in\cY$ and $\TTheta\in\cQ$, respectively. Fixed realizations of data and parameters are denoted by the lowercase letters $\yy$ and $\ttheta$. Bayesian inference requires specifying a joint probability measure for data and parameters on $\cY\times\cQ$, denoted $\pr_{\YY,\TTheta}$. This joint probability measure is typically specified as the product of the conditional probability measure of data $\YY$ given parameters $\TTheta = \ttheta$, denoted $\pr_{\YY|\ttheta}$, and the \emph{prior} probability measure of $\TTheta$, denoted $\pr_{\TTheta}$. Given the observed data $\YY = \yy$, let  $\pr_{\TTheta|\yy}$ be the \emph{posterior} probability measure. In addition, denote the \emph{marginal} probability measure for the data by $\pr_{\YY}$. 

In reality, joint models of parameters and data are typically characterized by their probability density functions. Formally, $\pr_{\YY|\ttheta}(d\yy) = f(\yy|\ttheta)\mu_{\cY}(dy)$, in which $f(\yy|\ttheta)$ is the \emph{likelihood} function and $\mu_{\cY}$ is a suitable dominating measure on the data space $\cY$. Similarly, $\pr_{\TTheta}(d\ttheta) = g(\ttheta)\mu_{\cQ}(d\ttheta)$, in which $g(\ttheta)$ is the \emph{prior density} and $\mu_{\cQ}$ is a dominating measure on the parameter space $\cQ$. Lebesgue measures and counting measures are typical choices of dominating measures when random variables are continuous and discrete, respectively. Correspondingly, the \emph{posterior density} that governs $\pr_{\TTheta|\yy}$ is proportional to $p(\ttheta, \yy) = f(\yy|\ttheta)g(\ttheta)$ in the light of Bayes' rule \citep[e.g.,][Section 1.3]{GelmanEtAl2013}. Without loss of generality, we focus on the case when $\cQ = \real^q$, a $q$-dimensional Euclidean space.

\subsection{Credible Regions and Posterior Possibility}

After observing $\yy$, Bayesian inference for model parameters can be made using a fundamental device: a posterior possibility contour. Intuitively, this contour function summarizes a family of nested credible regions across all credible levels $\alpha\in[0, 1]$. The contour not only facilitates the visualization and reconstruction of credible regions but also induces a possibility measure, a set function that maps any hypothesis (i.e., subset of the parameter space) to a value in the unit interval $[0, 1]$. Posterior possibilities are upper bounds of posterior probabilities, and hence warrant conservative Bayesian inference. We next provide a formal introduction to the above heuristics.

Let $C_\alpha(\yy)$ be a family of \emph{nested credible regions} indexed by $\alpha\in[0, 1]$ such that $\pr_{\TTheta|\yy}\{C_\alpha(\yy)\}\ge 1 - \alpha$ and that $C_\alpha(\yy)\subseteq C_{\alpha'}(\yy)$ whenever $\alpha\ge\alpha'$.\footnote{In practice, posterior distributions are often continuous; therefore, it is often possible to obtain nested credible regions with $\pr_{\TTheta|\yy}\{C_\alpha(\yy)\} = 1 - \alpha$ for every $\alpha\in(0, 1)$.} Nested credible regions $C_\alpha(\yy)$ can be conveniently constructed using a test statistic $T: \cY\times\cQ\to\real$ evaluated at the observed data $\yy$:
\begin{equation}
  C_{\alpha}(\yy) = \{\ttheta\in\cQ: T(\yy, \ttheta)\le\xi(\alpha)\},
  \label{eq:credtest}
\end{equation}
in which $\xi(\alpha) = \inf\{\xi\in\real:\pr_{\TTheta|\yy}\{T(\yy, \TTheta)\le\xi\}\ge 1 - \alpha\}$, the $(1 - \alpha)$th quantile of $T(\yy, \TTheta)$ under the posterior. Example credible regions include, but are not limited to, elliptical regions based on Laplace's approximation \cite[e.g.,][Section 13.3]{GelmanEtAl2013} and highest posterior density (HPD) regions \cite[e.g.,][Section 2.8]{BoxTiao2011}. Their corresponding test statistics will be provided in the Section ``A Practical Calibration Algorithm.''

\begin{figure}[!t]
  \centering
  \includegraphics[width=0.95\textwidth]{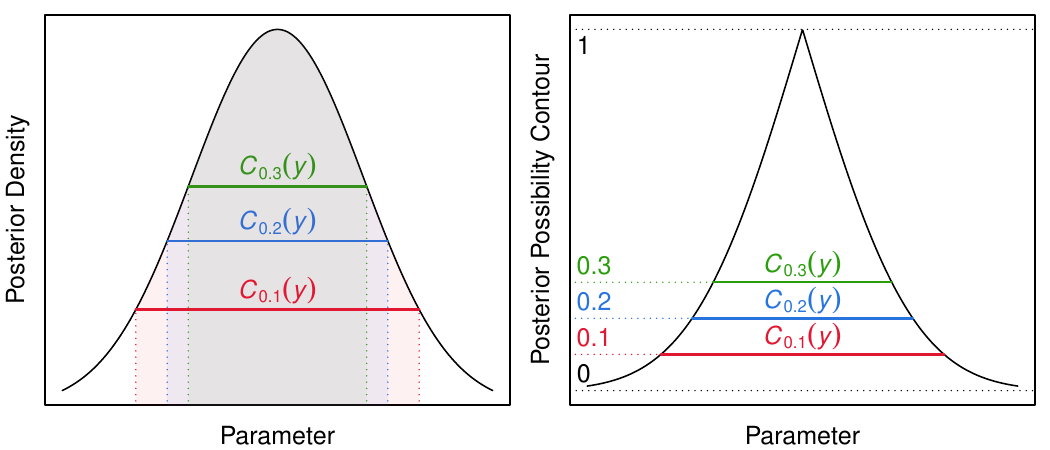}
  \caption{Graphical illustration for posterior density and possibility contour. Left: Thick colored horizontal line segments are credible regions $C_\alpha(\yy)$ for $\alpha = .1$ (red), $.2$ (blue), and $.3$ (green). They are regions with 90\%, 80\%, and 70\% posterior probabilities, depicted by the shaded area under the posterior density with matching colors. Right: The same three credible intervals are repositioned vertically to match their $\alpha$ levels. Stitching credible regions in the same fashion across all $\alpha$ levels yields the posterior possibility contour function.}
  \label{fig:denscont}
\end{figure}
Credible regions are more than just set estimators of model parameters. Their foundational role in Bayesian inference can be formally established by the construction of a \emph{posterior possibility contour} \citep{Zadeh1978, DuboisPrade1988, Dubois2006}. Comprehensive expositions of possibility theory and its applications in statistical inference can be found in, for example, \citet{DenoeuxLi2018}, \citet{LiuMartin2024}, and \cite{Martin2025b}. Let 
\begin{equation}
  \varpi_{\yy}(\ttheta) = \sup\{\alpha\in[0, 1]: \ttheta\in C_\alpha(\yy)\},
  \label{eq:postcont}
\end{equation}
the supremum of all $\alpha$-levels such that the parameter vector $\ttheta$ is contained in $C_\alpha(\yy)$. Because $\sup_{\ttheta\in\cQ}\varpi_{\yy}(\ttheta) = 1$, $\varpi_{\yy}$ is indeed a possibility contour.\footnote{Any function $h: \cX\to[0, 1]$ satisfying $\sup_{x\in\cX}h(x) = 1$ is commonly referred to as a possibility distribution in the literature of possibility calculus \cite[e.g.,][]{DuboisPrade1988}. We, however, call it a possibility contour following the IM convention \citep[e.g.,][]{LiuMartin2024} to avoid confusion with probability distributions.} (\ref{eq:postcont}) can be intuitively pictured as stitching together the family of nested credible regions at all $\alpha$ levels; a graphical illustration can be found in Figure \ref{fig:denscont}.  Conversely, we can extract credible intervals at any desired credibility level directly from the posterior possibility contour. Take the \emph{upper $\alpha$-level set} of the contour function $\varpi_\yy$, denoted by $\tilde C_\alpha(\yy) = \{\ttheta\in\cQ: \varpi_\yy(\ttheta) \ge \alpha\}$. Since $C_\alpha(\yy)\subseteq \tilde C_\alpha(\yy)$, the upper $\alpha$-level set constitutes a more conservatie credible region. In most applications, including all the examples in this paper, the family of nested credible regions of interest is left-continuous: that is, $C_\alpha(\yy) = \bigcap_{\alpha' < \alpha}C_{\alpha'}(\yy)$ for all $\alpha$. In this common scenario, the upper $\alpha$-level cut of the posterior possibility contour, $\tilde C_\alpha(\yy)$, coincides with the original $100(1 - \alpha)\%$ credible region, $C_\alpha(\yy)$.

By \citet[][Chapter 1]{MartinLiu2015}, an inferential procedure can be mathematically  represented as a set function, $2^{\cQ}\to[0, 1]$, which assigns a value in the unit interval to any subset of the parameter space. A posterior possibility contour $\varpi_\yy$ implies such a set function, termed a \emph{posterior possibility measure} $\overline\Pi_{\yy}$. Given a hypothesis $H\subseteq\cQ$, let the \emph{possibility} of $H$ be $\overline\Pi_{\yy}\{H\} = \sup_{\ttheta\in H}\varpi_{\yy}(\ttheta)$; for singleton hypotheses of form $H = \{\ttheta_0\}$, the definition reduces to $\overline\Pi_{\yy}\{\{\ttheta_0\}\} = \varpi_\yy(\ttheta_0)$. In words, the possibility of $H$ is defined as the supremum of the posterior possibility contour over $H$. This definition is intuitive in the sense that a hypothesis composed of multiple values must be no less plausible than any single value within it. An important property of the posterior possibility measure is \emph{compatibility}:
\begin{equation}
  \overline\Pi_{\yy}\{H\}\ge \pr_{\TTheta|\yy}\{H\}
  \label{eq:compatib}
\end{equation}
for any $\pr_{\TTheta|\yy}$-measurable hypothesis $H$ \citep{CousoEtAl2001}. To establish (\ref{eq:compatib}), note that $H\subseteq C_{\overline\Pi_{\yy}\{H\} + \varepsilon}(\yy)^c$ for all $\varepsilon > 0$. Because $C_{\overline\Pi_{\yy}\{H\} + \varepsilon}(\yy)^c$ is the complement of a $100(1 - \overline\Pi_{\yy}\{H\} - \varepsilon)\%$ credible region, $\pr_{\TTheta|\yy}\{H\}\le\pr_{\TTheta|\yy}\{C_{\overline\Pi_{\yy}\{H\} + \varepsilon}(\yy)^c\}\le\overline\Pi_{\yy}\{H\} + \varepsilon$, and (\ref{eq:compatib}) follows from sending $\varepsilon$ to 0. By (\ref{eq:compatib}), $\overline\Pi_{\yy}\{H\}$ can also be interpreted as an upper posterior probability of the hypothesis $H$.\footnote{The posterior possibility contour $\varpi_\yy$ also implies a necessity measure $\underline\Pi_\yy$, which is defined by duality as $\underline\Pi_\yy\{H\} = 1 - \overline\Pi_\yy\{H^c\}$ and pertains to a lower-probabilistic interpretation analogous to (\ref{eq:compatib}). The necessity measure, however, is not needed hereafter.}

Posterior possibilities are the cornerstone of inference within the Bayesian framework. The possibility contour function $\varpi_\yy$ provides a concise summary of nested credible regions: A set estimator for any prescribed $\alpha\in[0, 1]$ can be straightforwardly obtained by taking the contour's upper $\alpha$-level set. Additionally in Bayesian testing, the credibility of a null hypothesis is quantified conservatively by its posterior possibility. For instance, a simple hypothesis $H = \{\ttheta_0\}$ is rejected if its possibility $\overline\Pi_\yy\{H\} = \varpi_\yy(\ttheta_0)$ is smaller than the prescribed (small) $\alpha$ level.\footnote{\yl{The procedure described here is equivalent to determining whether $\ttheta_0$ is contained within a $100(1 - \alpha)\%$ credible region---a direct analogue to frequentist hypothesis testing. We are aware that many Bayesians prefer to use Bayes factors for hypothesis testing \citep[e.g.,][]{BergerDelampady1987}; however, we restrict our present discussion to possibility-based inference due to our focus on ``pragmatic Bayes.''}}

\subsection{Statistical Validity}
For any statistical procedure to be reliable in practice, it is desirable to establish that the procedure produces reliable results across varieties of scenarios. Specifically when making inference about model parameters, it is important not to repeatedly assign low possibility values to frequently occurring events in the long run. More and more Bayesian and frequentist statisticians, albeit holding different view on what probability represents, endorse the fundamental importance to evaluate the performance of inferential procedures over repeated samples \citep[e.g.,][]{Grunwald2018, Martin2022a, Martin2022b, Martin2022c}. Next, we review the notions of Bayesian and frequentist validity. We highlight that Bayesian inference based on a posterior possibility measure satisfies Bayesian validity, under the assumption that models for both parameters and data are correctly specified. Meanwhile, procedures with frequentist validity are automatically valid in the Bayesian sense for all priors, provided the data model is correctly specified.

\subsubsection{Bayesian Validity}
There are two forms of Bayesian validity: strong and weak. Intuitively, strong Bayesian validity requires that random parameters generated from the true prior distribution are rarely (i.e., with a long-run probability $\le\alpha$) assigned low possibilities (i.e., $\le\alpha$) over repeated generations of parameters and data. Meanwhile, weak Bayesian validity, implied by strong Bayesian validity, ensures that any hypothesis with a low possibility (i.e., $\le\alpha$) is unlikely (i.e., with a long-run probability $\le\alpha$) to contain the true random parameters. Importantly, inference based on posterior possibility measures (\ref{eq:postcont}) satisfies strong (and therefore weak) Bayesian validity, provided the prior is correctly specified. We formalize these definitions below.

Let $h:\cY\times \cQ\to\real$ be any $\pr_{\YY, \TTheta}$-integrable function. In the light of Fubini's Theorem \citep[][Theorem 18.3]{Billingsley2012}, we can write the joint expectation of $h$ with respect to data and parameters as the following iterated expectations:
\begin{equation}
  \iint h(\yy, \ttheta)\pr_{\YY, \TTheta}(d\yy, d\ttheta) = \int\left[\int h(\yy, \ttheta)\pr_{\TTheta|\yy}(d\ttheta) \right]\pr_{\YY}(d\yy).
  \label{eq:fubini}
\end{equation}
In (\ref{eq:fubini}), setting $h(\yy, \ttheta)$ to the indicator function of $\varpi_{\yy}(\ttheta)\le \alpha$ yields
\begin{equation}
  \pr_{\YY, \TTheta}\{\varpi_{\YY}(\TTheta)\le \alpha\} = \int\pr_{\TTheta|\yy}\{\varpi_{\yy}(\TTheta)\le \alpha\}\pr_{\YY}(d\yy) \le\alpha
  \label{eq:bcalib}
\end{equation}
for any $\alpha\in[0, 1]$, in which the last inequality follows from (\ref{eq:postcont}), the construction of $\varpi_\yy$. (\ref{eq:bcalib}) provides the definition of \emph{strong Bayesian validity} \citep{Martin2022b}, guaranteeing that assigning low possibilities to the true parameters is unlikely to occur in the long run.\footnote{\citet{Martin2022b} addressed a more general scenario when the prior is only partially specified.} Now let $H(\yy)\subseteq\cQ$ be a potentially data-dependent hypothesis about model parameters. (\ref{eq:bcalib}) further implies that
\begin{equation}
  \begin{aligned}
     \pr_{\YY, \TTheta}\{\TTheta\in H(\YY),\ \overline\Pi_{\YY}\{H(\YY)\}\le\alpha\} &=  \int\pr_{\TTheta|\yy}\{\TTheta\in H(\yy),\ \overline\Pi_{\yy}\{H(\yy)\}\le\alpha\}\pr_{\YY}(d\yy) \\
    &\le \int\pr_{\TTheta|\yy}\{\varpi_{\yy}(\TTheta)\le\alpha\}\pr_{\YY}(d\yy)\le\alpha,
  \end{aligned}
  \label{eq:bcalib1}
\end{equation}
in which the last line is due to the fact that $\varpi_{\yy}(\ttheta)\le\overline\Pi_{\yy}\{H(\yy)\}$ whenever $\ttheta\in H(\yy)$. (\ref{eq:bcalib1}) is referred to as \emph{weak Bayesian validity} in \citet{Martin2022b}. As a corollary of strong Bayesian validity, (\ref{eq:bcalib1}) guarantees that if a hypothesis is assigned a low possibility, it is unlikely to encompass the true parameters in the long run.\footnote{We implicitly assume that the events involved in (\ref{eq:bcalib}) and (\ref{eq:bcalib1}) are measurable.}

\subsubsection{Prior Misspecification and False Confidence}
Bayesian validity provides a justification for the performance of posterior-based inference over repeated sampling; nevertheless, it hinges upon the crucial assumption of correct prior specification. Stated differently, the same prior measure $\pr_{\TTheta}$ must be involved in both data generation (i.e., forming $\pr_{\YY, \TTheta}$) and statistical inference (i.e., forming $\pr_{\TTheta|\yy}$). When this assumption is violated, the validity guarantee may fail and erroneous inference may result. One such example is the False Confidence Theorem \citep[FCT;][]{BalchEtAl2019}. When the true prior is a point mass (i.e., the classical frequentist setup with fixed true parameters), any non-degenerate prior is misspecified. In this case, we can always find false hypotheses that are frequently assigned high posterior possibilities.

Let the true parameter-generating prior be $\tilde\pr_{\TTheta} \ne \pr_{\TTheta}$, then the previous iterated expectation (\ref{eq:fubini}) no longer holds true but should be modified to
\begin{equation}
  \iint h(\yy, \ttheta)\tilde\pr_{\YY, \TTheta}(d\yy, d\ttheta) =  \int\left[\int h(\yy, \ttheta)\frac{d\tilde\pr_{\TTheta|\yy}}{d\pr_{\TTheta|\yy}}(\ttheta)\pr_{\TTheta|\yy}(d\ttheta) \right]\tilde\pr_{\YY}(d\yy).
  \label{eq:fubini1}
\end{equation}
In (\ref{eq:fubini1}), the incorrect posterior $\pr_{\TTheta|\yy}$ is deduced from the incorrect prior $\pr_{\TTheta}$, while the correct posterior $\tilde\pr_{\TTheta|\yy}$ and marginal $\tilde\pr_{\YY}$ are obtained from the correct $\tilde\pr_{\TTheta}$. $d\tilde\pr_{\TTheta|\yy}/d\pr_{\TTheta|\yy}$ denotes the Radon-Nikodym derivative (e.g., a density ratio, essentially) of the correct posterior with respect to incorrect posterior, which is assumed to exist. Because the right-hand side of (\ref{eq:fubini1}) differs from (\ref{eq:fubini}), Bayesian validity (\ref{eq:bcalib}) and (\ref{eq:bcalib1}) are no longer satisfied in general. 

A direct corollary of the FCT provides a prominent example of invalidity: when the true parameters are fixed (i.e., the true prior $\tilde\pr_{\TTheta}$ is a point mass concentrated at some $\ttheta_0$), no Bayesian inference derived from a non-degenerate prior can guarantee weak validity (\ref{eq:bcalib1}). For any $\alpha\in[0, 1]$, the FCT implies the existence of a hypothesis $C_\alpha\subseteq\cQ$ such that $\ttheta_0\notin C_\alpha$ but $C_\alpha$ is a $100(1 - \alpha)\%$ credible region with arbitrarily large $\pr_{\YY|\ttheta_0}$-probability.\footnote{In the proof of the FCT, $C_\alpha$ can be constructed as the complement of a ball concentrated around $\ttheta_0$.} Then weak Bayesian validity (\ref{eq:bcalib1}) fails if we define $H(\yy)\equiv C_\alpha^c$---ensuring that $\ttheta_0\in H(\yy)$ always holds true---and construct a possibility measure $\overline\Pi_\yy$ via (\ref{eq:postcont}) by setting $C_\alpha(\yy) = C_\alpha$ whenever $\pr_{\TTheta|\yy}\{C_\alpha\}\ge 1 - \alpha$.

\subsubsection{Frequentist Validity}
In practice, we often cannot evaluate the extent to which the parameter-generating prior deviates from the chosen prior for inference, and consequently, the degree to which Bayesian inference is vulnerable to false confidence. In complete ignorance of the parameter-generating mechanism, a safer alternative is to rely on procedures that are valid in the frequentist sense, because it necessarily implies Bayesian validity with any parameter-generating priors. Intuitively, frequentist validity requires that the possibility measure used for inference should not frequently (i.e., with long-run probability $\le\alpha$) assign a low possibility (i.e., $\le\alpha$) to any fixed true data-generating parameters, uniformly across the entire parameter space. We next formally define this notion. 

Let $\pi_{\yy}:\cQ\to[0, 1]$ be a general possibility contour function that satisfies $\sup_{\ttheta\in\cQ}\pi_{\yy}(\ttheta) = 1$. $\pi_{\yy}$ is not necessarily derived from any posterior. The \emph{frequentist validity} requires that 
\begin{equation}
  \sup_{\ttheta\in\cQ}\pr_{\YY|\ttheta}\{\pi_\YY(\ttheta)\le\alpha\}\le\alpha
  \label{eq:fcalib}
\end{equation}
for all $\alpha\in[0, 1]$. An important family of possibility contour functions that satisfy (\ref{eq:fcalib}) consists of the survival function of a test statistic $T(\YY, \ttheta)$ under $\pr_{\YY|\ttheta}$ evaluated at its observed value $T(\yy, \ttheta)$:
\begin{equation}
  \pi_\yy(\ttheta) = \pr_{\YY|\ttheta}\{T(\YY, \ttheta) \ge T(\yy, \ttheta)\}.
  \label{eq:pval}
\end{equation}
Resulting from the probability integral transform, $\pi_{\yy}$ defined by (\ref{eq:pval}) satisfies (\ref{eq:fcalib}) \citep[e.g.,][Theorem 2.1.10 and Exercise 2.10]{CasellaBerger2002}. \yl{Technically, (9) will correspond to a possibility contour so long as $\inf_{\ttheta \in \cQ} T(\yy,\ttheta)$ does not depend on $\yy$. Fortunately, this additional requirement is satisfied when the test statistic is non-negative and attains zeros at some $\ttheta$ for each $\yy$; this is true, for example, of the Wald and posterior density ratio statistics we consider in this paper.} Note that constructions of the form (\ref{eq:pval}) are also commonly referred to as \emph{$p$-value functions} \citep{Fraser2019, MartinLiu2014, SchwederHjort2016, XieSingh2013}.

The frequentist validity (\ref{eq:fcalib}) implies the strong Bayesian validity (\ref{eq:bcalib}) for all generating mechanism $\tilde\pr_{\YY, \TTheta}$ that composes $\tilde\pr_{\TTheta}$ and $\pr_{\YY|\ttheta}$:
\begin{equation}
  \tilde\pr_{\YY, \TTheta}\{\pi_{\YY}(\TTheta)\le \alpha\} = \int\pr_{\YY|\ttheta}\{\pi_{\YY}(\TTheta)\le \alpha\}\tilde\pr_{\TTheta}(d\ttheta) \le\sup_{\ttheta\in\cQ}\pr_{\YY|\ttheta}\{\pi_\YY(\ttheta)\le\alpha\}\le\alpha.
  \label{eq:strongcalib}
\end{equation}
Under the assumption that the data model $\pr_{\YY|\ttheta}$ is correctly specified, a major implication of (\ref{eq:strongcalib}) is that any valid inferential procedure in the frequentist sense is automatically valid in the Bayesian sense regardless of the parameter-generating prior $\tilde\pr_{\TTheta}$. In contrast, inferential procedures derived directly from a Bayesian posterior are not necessarily valid if the specified prior disagrees with the parameter-generating prior. While it is feasible to evaluate model misspecification via goodness-of-fit diagnostics (i.e., correctness of $\pr_{\YY|\ttheta}$), there often lack reliable ways to assess prior misspecification except in MC experiments or in contexts of research syntheses. We therefore recommend that Bayesian methods, whenever used in a ``pragmatic'' mode, should be calibrated to achieve frequentist validity.

\section{Calibrating Bayesian Inference}
In this section, we present a generic computational strategy to calibrate posterior possibility contours and ensure frequentist validity (\ref{eq:fcalib}). As input, we define a family of nested credible regions by thresholding an observed test statistic. We then use a gradient-free SA algorithm to calibrate the credible regions based on the test statistic's survival function. Notably, the optimization program we solve is essentially equivalent to finding a ``variational-like approximation'' to the IM possibility contour (\ref{eq:pval}) \citep{CellaMartin2024, Martin2025a}. A unique contribution of our proposal is the novel combination of simultaneous perturbation SA \citep{Spall1992, Spall2000, Spall2009} and manifold optimization \citep{AbsilMalick2012, AbsilEtAl2008}, which enhances the scalability of the calibration algorithm and facilitates its applications in models of realistic sizes. Moreover, we address not only simultaneous inference for all parameters but also the marginal inference for a single focal parameter.

\subsection{Test Statistics and Nested Credible Regions}

Specifically in the current work, we construct nested credible regions from three types of test statistics. For simultaneous inference of all model parameters, we rely on the Wald statistic and the posterior density ratio (PDR) statistic, which generate elliptical and HPD credible regions, respectively. For the marginal inference of a focal parameter, we use the marginal Wald statistic to obtain symmetric credible intervals.

\subsubsection{Wald Statistic and Elliptical Regions}

Let $\hat\ttheta(\yy) = \argmax_{\ttheta\in\real^q} p(\ttheta, \yy)$ be the maximum \emph{a posteriori} (MAP) estimator of the model parameters; it is assumed that the MAP estimator uniquely exists for all $\yy\in\cY$. We define the \emph{Wald statistic} as the quadratic form:
\begin{equation}
  T_{\rm W}(\yy, \ttheta) = \left[\hat\ttheta(\yy) -  \ttheta\right]\t\hat\SSigma_{\ttheta}(\yy)^{-1}\left[\hat\ttheta(\yy) -  \ttheta\right],
  \label{eq:wald}
\end{equation}
in which $\hat\SSigma_{\ttheta}(\yy)$ is an estimated covariance matrix of the MAP estimator. Choices of $\hat\SSigma_{\ttheta}(\yy)$ include, but are not limited to, the inverse minus expected Hessian of the log-posterior $-\ex_{\YY|\ttheta}[\nabla_{\ttheta\ttheta\t}^2\log p(\ttheta, \yy)]^{-1}$ or its sample counterpart, provided these matrices are non-negative definite. The family of nested credible region associated with the Wald statistic can be expressed as
\begin{equation}
  D^{\rm W}_{\xi}(\yy) = \{\ttheta\in\real^q: T_{\rm W}(\yy, \ttheta)\le \xi\}.
  \label{eq:ellip}
\end{equation}
For notational convenience, we now index the family of credible intervals in (\ref{eq:ellip}) by $\xi\ge 0$, the threshold of the observed test statistic. To link it back to the more common indexing using the credibility level $\alpha\in[0, 1]$, let $\alpha(\xi) =  1 - \pr_{\TTheta|\yy}\{D_\xi^W(\yy)\}$ and note that $D_\xi^{\rm W}(\yy)$ is a $100[1 - \alpha(\xi)]\%$ credible region, previously denoted by $C_{\alpha(\xi)}(\yy)$. Geometrically, (\ref{eq:wald}) is an elliptical region because $T_{\rm W}$ is a non-negative quadratic form in $\ttheta$.

\subsubsection{Posterior Density Ratio Statistic and Highest Posterior Density Regions}
Alternatively, we can define credible regions based on the PDR statistic:
\begin{equation}
  T_{\rm PDR}(\yy, \ttheta) = -2\left[\log p(\ttheta, \yy) - \log p\left(\hat\ttheta(\yy)\,|\,\yy \right) \right].
  \label{eq:pdr}
\end{equation}
(\ref{eq:pdr}) is a generalization of the likelihood-ratio statistic in standard large-sample theory of maximum likelihood estimation. Also note that the PDR statistic (\ref{eq:pdr}) is a logarithmic transform of the ``relative plausibility ordering'' defined by \citet{Martin2022b} in a more general context concerning partial priors. Similar to (\ref{eq:ellip}), credible regions can be constructed by collecting parameter values with sufficiently low PDR statistic values:
\begin{equation}
  D^{\rm PDR}_\xi(\yy) = \{\ttheta\in\real^q: T_{\rm PDR}(\yy, \ttheta)\le \xi\},
  \label{eq:hpd}
\end{equation}
where $\xi\ge 0$.  Due to the one-to-one correspondence between the PDR statistic (\ref{eq:pdr}) and the posterior density $p(\ttheta, \yy)$, the credible region defined by (\ref{eq:hpd}) amounts to the HPD region.

\subsubsection{Marginal Wald Statistic and Symmetric Intervals}

Define a partition of the parameter vector $\ttheta = (\varphi, \nnu\t)\t$, in which $\varphi\in\real$ denotes the \emph{focal parameter} and $\nnu$ collects the remaining \emph{nuisance parameters}. To make inference about $\varphi$, we define the corresponding marginal Wald statistic by 
\begin{equation}
  T_{\varphi}(\yy, \varphi) = \frac{\left[\hat\varphi(\yy) - \varphi\right]^2}{\hat\sigma_{\varphi}^2(\yy)},
  \label{eq:mwald}
\end{equation}
in which $\hat\sigma^2_\varphi(\yy)$ is the first diagonal entry of $\hat\SSigma_{\ttheta}(\yy)$. The family of credible regions constructed from (\ref{eq:mwald}) can be equivalently represented by
\begin{equation}
  D_\xi^\varphi(\yy) = \{\ttheta\in\real^q: T_{\varphi}(\yy, \varphi)\le\xi\},
  \label{eq:int}
\end{equation}
where $\xi\ge 0$.  The region defined by (\ref{eq:int}) a cylinder set in the parameter space $\real^q$. This is because the test statistic is invariant to any changes in $\nnu$. Projecting (\ref{eq:int}) onto the focal parameter space yields a credible interval that is symmetric around $\hat\varphi(\yy)$.

Under further regularity conditions of the Bernstein von-Mises theorem \citep[e.g.,][Section 5.5]{BickelDoksum2015}, the Wald statistic (\ref{eq:wald}), the PDR statistic (\ref{eq:pdr}), and the marginal Wald statistic (\ref{eq:mwald}) are all asymptotically chi-square when evaluated at the (fixed) true parameters. The chi-square approximations, however, can be inaccurate in finite samples, potentially leading to invalid inference (see the ``Monte Carlo Experiment'' section for an empirical evaluation). We next discuss a practical SA algorithm that can be used to calibrate the credible regions (\ref{eq:ellip}), (\ref{eq:hpd}), and(\ref{eq:int})  to achieve frequentist validity.

\subsection{Calibration by Simultaneous-Perturbation Riemannian Stochastic Approximation}

\subsubsection{Calibration Problem}

To achieve frequentist validity, we can calibrate credible regions generated by the test statistic $T$ using its own $p$-value function. Let $\pi_\yy(\ttheta)$ be the $p$-value function (\ref{eq:pval}) of the test statistic $T$, which defines the credible region $D_\xi(\yy) = \{\ttheta\in\real^q: T(\yy, \ttheta)\le \xi\}$. For each threshold $\xi\ge 0$, calibration amounts to finding
\begin{equation}
  \alpha_{\yy}^*(\xi) = 
  \sup_{\ttheta\in\partial D_{\xi}(\yy)}\pi_{\yy}(\ttheta),
  \label{eq:optim}
\end{equation}
in which $\partial D_\xi(\yy) = \{\ttheta\in\cQ: T(\yy, \ttheta) = \xi\}$ is the boundary of $D_\xi(\yy)$. We term (\ref{eq:optim}) the \emph{calibrated $\alpha$ level} at the threshold value $\xi$. For any parameter vector $\ttheta\in\cQ$, the corresponding \emph{calibrated posterior possibility contour} is defined as
\begin{equation}
  \tilde\varpi_{\yy}(\ttheta) = \alpha_{\yy}^*(T(\yy, \ttheta)).
  \label{eq:calibpi}
\end{equation}
Because $\ttheta$ itself is a member of $\partial D_{T(\yy, \ttheta)}(\yy)$, we have
\begin{equation}
  \sup_{\ttheta\in\cQ}\pr_{\YY|\ttheta}\{\tilde\varpi_{\YY}(\ttheta)\le\alpha\}\le \sup_{\ttheta\in\cQ}\pr_{\YY|\ttheta}\{\pi_{\YY}(\ttheta)\le\alpha\}\le \alpha 
  \label{eq:ccalib}
\end{equation}
 for all $\alpha\in[0, 1]$, establishing frequentist validity (\ref{eq:fcalib}).

 We make further assumptions to ensure that the constrained program on the right-hand side of (\ref{eq:optim}) is sufficiently regular. First, the feasible region is a $\xi$-level set of the test statistic (viewed as a function of $\ttheta$), which is difficult to characterize without further restrictions. To enable a neat geometric characterization of the level set, we assume that the test statistic is differentiable in $\ttheta$ and that $\xi$ is a regular value of the statistic. Under these additional assumptions, $\partial D_\xi(\yy)$ is a differential submanifold of the $q$-dimensional Euclidean space.\footnote{If the Jacobian $\nabla_{\ttheta}T(\yy, \ttheta)\in\real^q$ does not vanish for all $\ttheta\in\partial D_\xi(\yy)$, then $\xi$ is regular.} Second, we assume that the $p$-value function $\pi_{\yy}(\ttheta)$ is differentiable for all $\yy\in\cY$ and $\ttheta\in\real^q$. 

\subsubsection{Algorithm}

To solve the optimization problem (\ref{eq:optim}), we propose a simultaneous-perturbation Riemannian stochastic approximation (SPRSA) algorithm. For clarity and accessibility, we provide only a heuristic overview of the algorithm in the main text, drawing analogies to the standard gradient ascent method. The pseudocode and further details of the SPRSA algorithm can be found in Appendix \ref{a:alg}. A review of geometric concepts related to embedded submanifolds and a convergence proof of the SPRSA algorithm are provided in the Supplementary Materials.

At the $k$th iteration ($k = 1, 2, \dots$) of the SPRSA algorithm, the parameter vector $\ttheta^{(k)}$ is updated as follows:
  \begin{equation}
    \ttheta^{(k + 1)} = \RR\left(\ttheta^{(k)}, a_k\MM(\ttheta^{(k)})\widehat{\nabla_{\ttheta}\pi_{\yy}}(\ttheta^{(k)})\right).
    \label{eq:sprsak}
  \end{equation}
  In (\ref{eq:sprsak}), $a_k > 0$ is the iteration-specific \emph{learning rate}, $\MM(\ttheta)$ is a $q\times q$ matrix-valued function of $\ttheta$, $\RR: \real^q\times\real^q\to\partial D_\xi(\yy)$ is a suitable map ensuring that the updated iterate remains on the manifold $\partial D_\xi(\yy)$, and $\widehat{\nabla_{\ttheta}\pi_{\yy}}$ is an estimated gradient of the $p$-value function. Indeed, if $\MM$ were the identity matrix, $\RR(\ttheta, \hh) = \ttheta + \hh$ for $\ttheta,\hh\in\real^q$, and the gradient $\nabla_{\ttheta}\pi_\yy$ could be exactly evaluated, then (\ref{eq:sprsak}) would reduce to the standard gradient ascent in $\real^q$.\footnote{We consider gradient ascent, not descent, because we aim to maximize the objective function (\ref{eq:optim}).} The general SPRSA algorithm can be applied to problems subject to manifold constraints and objective functions whose gradient cannot be computed exactly. These are precisely the challenges in finding the calibrated $\alpha$ level (\ref{eq:optim}).

  Optimization problems whose feasible regions form differentiable submanifolds of Euclidean spaces can be efficiently solved by Riemannian gradient algorithms (\citealt{AbsilEtAl2008}; see also \citealt{Liu2020,Liu2021} for accessible introductions to Riemannian gradient algorithms and their applications in psychometric problems). A Riemannian gradient ascent algorithm employs specific choices of $\MM$ and $\RR$ in (\ref{eq:sprsak}) that account for the local geometry of the manifold. First, the steepest ascent direction on a differentiable submanifold is defined locally by the \emph{Riemannian gradient}, which is obtained by projecting the ambient gradient of the objective function onto the tangent space of the submanifold. Therefore, the matrix-valued function $\MM(\ttheta)$ in (\ref{eq:sprsak}) is set to the corresponding projection matrix. The exact expression of $\MM(\ttheta)$ is provided in Appendix \ref{a:alg}. Second, a step taken along the Riemannian gradient direction leaves the manifold and must therefore be pulled back onto it via the \emph{retraction} map $\RR$. Specifically, $\RR(\ttheta, \hh)$ takes as input the current iterate $\ttheta$ on the manifold and a tangent vector $\hh$ and then returns an updated iterate on the manifold; it preserves the direction of $\hh$ in the vicinity of $\ttheta$. The exact definition of the retraction map can be found in Appendix \ref{a:alg}.

  A further challenge in applying the standard Riemannian gradient ascent algorithm to solve (\ref{eq:optim}) arises when evaluating the exact gradient of the $p$-value function, $\nabla_{\ttheta}\pi_{\yy}(\ttheta)$. This is because the $p$-value function is an integral whose domain depends on $\ttheta$ through the test statistic. Although analytical derivatives are sometimes available via the generalized Leibniz rule \citep[also known as the Reynolds Transport Theorem; see, e.g.,][]{Flanders1973, ReddigerPoirier2023}, evaluating them is generally difficult because the computation of the test statistic itself may involve optimization. In the SA literature, a well-documented strategy for handling intractable gradients is to approximate them by finite differences \citep[FDs; e.g.,][]{KieferWolfowitz1952}. In particular, we consider a simultaneous-perturbation FD approximation of the gradient due to \citet{Spall1992}, in which a random perturbation is introduced to all parameters simultaneously, and the magnitude of perturbation decays along iterations at a suitable rate. The formula of the SP approximation and the conditions on the decay rate are presented in Appendix \ref{a:alg}.

\section{Monte Carlo Experiment}

\yl{In this section, we numerically compare the performance of calibrated versus standard Bayesian inference in two models: Gaussian location-scale regression and Gaussian linear factor analysis.} Calibrated posterior possibilities are computed via the proposed SPRSA algorithm, while the standard posterior possibilities are obtained using both asymptotic (chi-square) approximation and MCMC sampling. \yl{We demonstrate that, under certain specifications of the data model and prior, uncalibrated Bayesian inference can be unacceptably liberal in small-sample problems.} In contrast, the proposed calibration procedure effectively generates valid inference across almost all simulated conditions.

\subsection{Study 1: Gaussian Location-Scale Regression}
\subsubsection{Data Generation}

Location-scale regression is an extension of linear regression accommodating heteroscedastic error terms \citep{Aitkin1987,Harvey1976,Verbyla1993}. The model allows not only the mean but also the variance of an outcome variable $Y_i\in\real$ to depend on fixed design variables $\xx_i\in\real^m$. Here we consider a parametric version of the model assuming normality: That is, we assume that $Y_i|\xx_i\sim\sN\left(\mu(\xx_i), \sigma^2(\xx_i)\right)$ for each $i = 1,\dots, n$, in which
\begin{equation}
  \mu(\xx_i) = \xx_i\t\bbeta\text{ and }\log\sigma(\xx_i) = \xx_i\t\ggamma
  \label{eq:locscale}
\end{equation}
are referred to as location and log-scale functions, respectively. The parameter vector $\ttheta = (\bbeta\t, \ggamma\t)\t\in\real^{q}$, where $q = 2m$, concatenates the location regression coefficients $\bbeta$ and the log-scale regression coefficients $\ggamma$. For each observation $i$, the log-likelihood function can be expressed (up to an additive constant) as
\begin{equation}
  \log f(y_i, \ttheta) = - \xx_i\t\ggamma - \frac{(y_i - \xx_i\t\bbeta)^2}{2\exp(\xx_i\t\ggamma)}.
  \label{eq:liklsr}
\end{equation}
To simulate the response for each $i$, a convenient data-generating algorithm is $Y_i = \mu(\xx_i) + \sigma(\xx_i)U_i$, where $U_i\sim\sN(0, 1)$. Location-scale regression is a special case of more general distributional regression frameworks such as the generalized additive models for location, scale and shape (GAMLSS; \citealt{RigbyStasinopoulos2005}) and the vector generalized additive models (VGAM; \citealt{Yee2015}).
Simulation conditions were determined by two crossed factors: number of design variables ($m = 3$ and 10), and three parameter-generating scenarios. The sample size was fixed at $n = 100$. For each observation $i$, the design variable vector $\xx_i$ was constructed by $\xx_i = (1, \tilde\xx_i\t)\t$, in which $\tilde\xx_i\in\real^{m - 1}$ consists of multivariate normal variates with a uniform pairwise correlation of .3.\footnote{To clarify, while design variables were randomly generated across replications, they were held constant during calibration in any single replication.} In Scenario 1, the location regression coefficients (i.e., $\bbeta$) were randomly sampled from $\sN(0, 1)$, while the log-scale regression coefficients (i.e., $\ggamma$) were sampled from $\sN(0, .2^2)$. This scenario corresponds to a Bayesian setup with a non-degenerate parameter-generating prior. In contrast, Scenario 2 featured a point-mass prior: the location coefficients were fixed at 1 and the log-scale coefficients were fixed at .2. In Scenario 3, all coefficients were randomly generated from $t_5(0, .5^2)$: a Student $t$ distribution with a location of 0, a scale of .5, and 5 degrees of freedom. The parameter-generating distribution in Scenario 3 matches our strongly informative prior specification, rendering the latter correctly specified. We used MATLAB version 25.2 \citep{MATLAB} to generate data; 512 replications were run under each condition. 

\subsubsection{Tuning Details}
We specified independent Student $t$ priors for all parameters, adopting a common practice for regression-type models \citep[e.g.,][Chapter 16]{GelmanEtAl2013}. Two prior specifications were considered: the strongly informative $t_5(0, .5^2)$ and the weakly informative $t_5(0, 25^2)$.  MCMC sampling was performed using JAGS \citep{JAGS}. In each replication, we ran 5 chains in parallel. The number of adaptation, burn-in, and retained iterations for each chain are 1000, 10000, and 10000, respectively. MC samples of parameters were stored at a thinning interval of 5 using the retained iterations from the 5 chains, totaled up to 10000 draws. The potential scale reduction factor (PSRF) and the effective sample size (ESS) for each coordinate of $\ttheta$ were recorded in each replication. We deemed the replication convergent if PSRF $\le$ 1.1 and ESS $\ge$ 100 for all parameters.

To evaluate posterior possibilities and perform calibration, we computed the Wald test statistics (\ref{eq:wald}), the PDR statistic (\ref{eq:pdr}), and the marginal Wald statistics (\ref{eq:mwald}) for $\beta_1$, $\beta_2$, $\gamma_1$, and $\gamma_2$, respectively. When evaluated at the true parameter values, both the Wald and PDR statistics approximately follow a $\chi^2_{q}$ distribution, while the marginal Wald statistics approximately follow a $\chi^2_1$ distribution. The MATLAB function \texttt{fminunc} with its default configuration was used to maximize the log-posterior of $\ttheta$. In the definition of the Wald test statistic, the covariance matrix of the MAP estimator was approximated by the observed Fisher information. 

We made a simplification when calibrating the possibility contour based on the marginal Wald statistic. The boundary of the marginal Wald credible interval (\ref{eq:mwald}), $\partial D_\xi^\varphi(\yy) = \{\ttheta: [\hat\varphi(\yy) - \varphi]^2 =\xi\hat\sigma_\varphi(\yy)^2\}$ is a union of two disjoint lines: $\{\ttheta: \varphi = \hat\varphi(\yy) \pm \sqrt{\xi}\hat\sigma_\varphi(\yy)\}$. Solving the literal calibration program (\ref{eq:optim}) requires performing numerical search on the two lines separately and taking the maximum of the two solutions. When evaluating at $\xi = T_{\varphi}(\yy, \phi)$ for any $\phi\in\real$, the two lines reduces to $\{\ttheta: \varphi = \phi\}$, which contains $\phi$, and its mirror image around $\hat\varphi(\yy)$, which does not contain $\phi$. Consequently, we only run the SPRSA algorithm on the first line. This approach not only halves the computational cost but also yields less conservative solutions.

Pilot simulations were conducted to tune the SPRSA algorithm. Specifically, the learning rate sequence was determined by $a_k = \underline a k^{-\underline b}$, where $\underline b = .651$. We set the initial learning rate $\underline a = 1$ for simultaneous inference (i.e., using the Wald and PDR statistics) and .01 for marginal inference (i.e., using marginal Wald statistics). The FD rate sequence was computed by $c_k = \underline c k^{-\underline d}$, in which $\underline c = .5$ and $\underline d = .15$. It can be straightforwardly verified that these two rate sequences satisfy the condition (\ref{eq:seq}). The number of iterations of the algorithm was set to $K = 50000$. The FD perturbation at the last iteration is then $.05\times 50000^{-.15}\approx .01$. The final solution was computed by the recursive averaging procedure (\ref{eq:avgest}) after a burn-in period of the first 10000 iterations. Calibrated $\alpha$-levels were then computed via 10000 additional simulations, holding the parameter values at the final averaged solution. We implemented the SPRSA algorithm in MATLAB. The simulation code is available at \href{https://github.com/yliu87/CalibBayes}{https://github.com/yliu87/CalibBayes}.

\subsubsection{Evaluation Criteria}

We evaluated the validity of posterior-based inference across three candidate methods: the chi-square approximation to the original posterior contour, the MCMC approximation to the original posterior contour, and the calibrated contour. In each replication, we used the observed data $\yy$ and the data-generating parameters $\ttheta\sim\pr_{\TTheta}^*$ to compute the chi-square and MCMC approximations to $\varpi_{\yy}(\ttheta)$, as well as the calibrated contour $\tilde\varpi_{\yy}(\ttheta)$, for each test statistic. Because the MCMC sampler does not always converge (meeting the criteria PSRF $\le$ 1.1 and ESS $\ge$ 100), we reported the convergence rate for MCMC sampling for each condition. We then graphed empirical distribution functions (EDFs) for the contour function values across replications under each simulated condition. An EDF curve below the diagonal line indicates conservative and thus valid inference, whereas a curve above the diagonal suggests the validity requirement has not been met. To account for MC error, comparisons with the diagonal line are benchmarked against its 95\% normal-approximation MC confidence band (i.e,. $\alpha\pm1.96\sqrt{\alpha(1-\alpha)/512}$). 

\subsubsection{Results}

\begin{table}[!t]
  \begin{center}
  \caption{Number of replications where Markov chain Monte Carlo sampling fails to converge.}
  \label{tab:cvg}
  \begin{tabular}{ccrrr}
   \toprule 
   \multirow{2}{*}{$m$} & 
   \multirow{2}{*}{Prior} & 
   \multicolumn{3}{c}{Scenarios}\\
   \cmidrule(r){3-5}
   & & 1 & 2 & 3\\
   \midrule 
   3 & Strong & 0 & 0 & 2\\
     & Weak & 0 & 0 & 2\\
     \midrule
   10 & Strong & 1 & 8 & 142\\
     & Weak & 1 & 10 & 150 \\
   \bottomrule
  \end{tabular}
  \end{center}
  \textit{Note. Convergence is declared if the potential scale reduction factor is less than 1.1 and the effective sample size is at least 100. $m$: Number of design variables.}
\end{table}

Table \ref{tab:cvg} summarizes convergence behavior of MCMC sampling across all combinations of parameter-generating scenarios and prior configurations. We observed that non-convergence occurs more frequently when the number of design variables is large ($m = 10$) and when the parameter-generating distribution is heavy-tailed (Scenario 3). The worst case scenario is when the weakly informative prior is used under Scenario 3, in which the MCMC sampler fails to converge in 150 out of 512 replications. In what follows, MCMC results are reported based solely on converged replications, whereas results for chi-square approximation and calibration are obtained using all replications. \yl{This missingness handling approach in Monte Carlo experiments was referred to as ``method-wise deletion'' by \citet{PawelEtAl2026}. As a robustness check, we also presented the results for $m = 10$ (where non-convergence is frequent) in the Supplementary Materials using ``repetition-wise deletion'', restricting the summary to replications where MCMC sampling converged.}

\begin{figure}[!t]
  \centering
  \includegraphics[width=\textwidth]{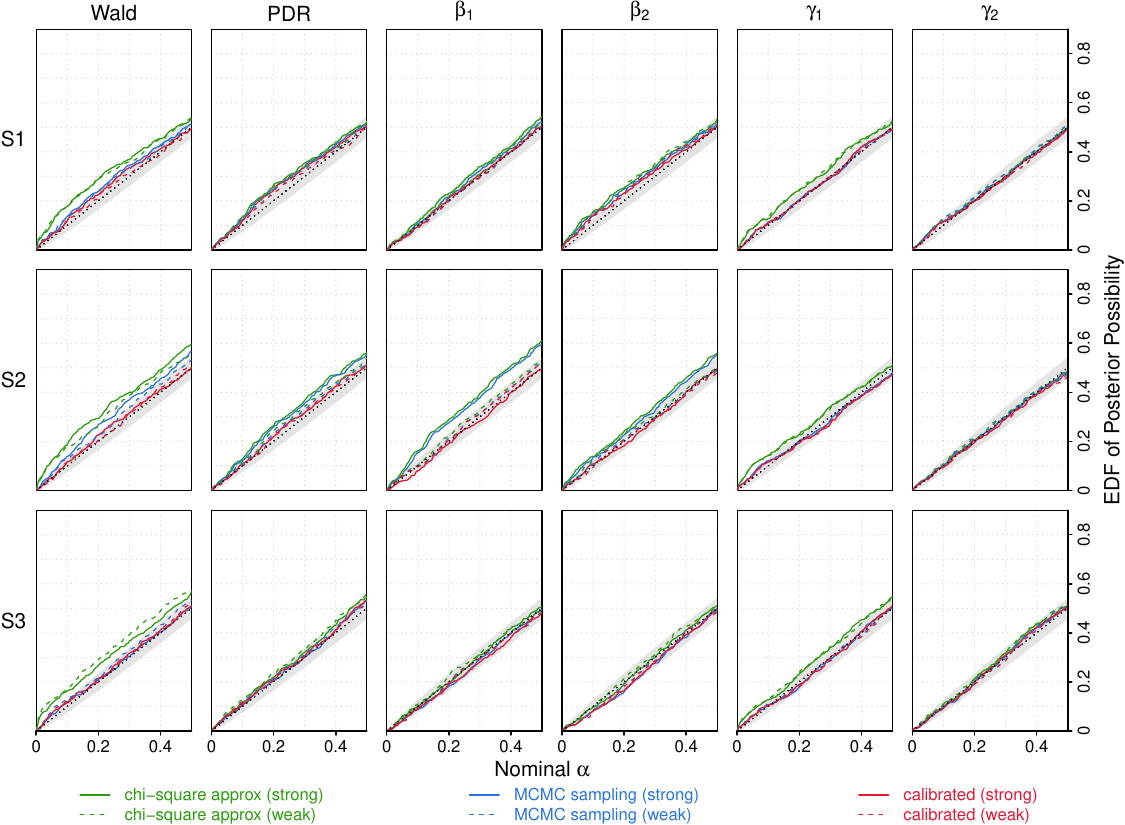}
  \caption{Summary of Study 1: $m = 3$ design variables. Rows of the graphical table represent three parameter-generating scenarios (S1--S3). Columns represent six types of test statistics: the first two columns correspond to the Wald and posterior density ratio (PDR) statistics for simultaneous inference of all parameters, and the remaining four columns correspond to the marginal Wald statistics for selected parameters ($\beta_1$, $\beta_2$, $\gamma_1$, and $\gamma_2$). Six empirical distribution functions (EDFs) of posterior possibilities are presented in each panel. Colors are used to contrast results based on chi-square approximation (green), Markov chain Monte Carlo (MCMC) sampling (blue), and the proposed calibration algorithm (red). Line types are used to distinguish strong ($t_5(0, .5^2)$; solid) and weak ($t_5(0, 25^2)$; dashed) priors. The diagonal dotted lines in each panel indicates exact uniformity; a 95\% normal-approximation, pointwise Monte Carlo confidence band is shown by the gray area. EDFs above the diagonal signifies liberal and thus invalid inference, while EDFs below the diagonal implies conservative and thus valid inference.}
  \label{fig:summ1}
\end{figure}

\begin{figure}[!t]
  \centering
  \includegraphics[width=\textwidth]{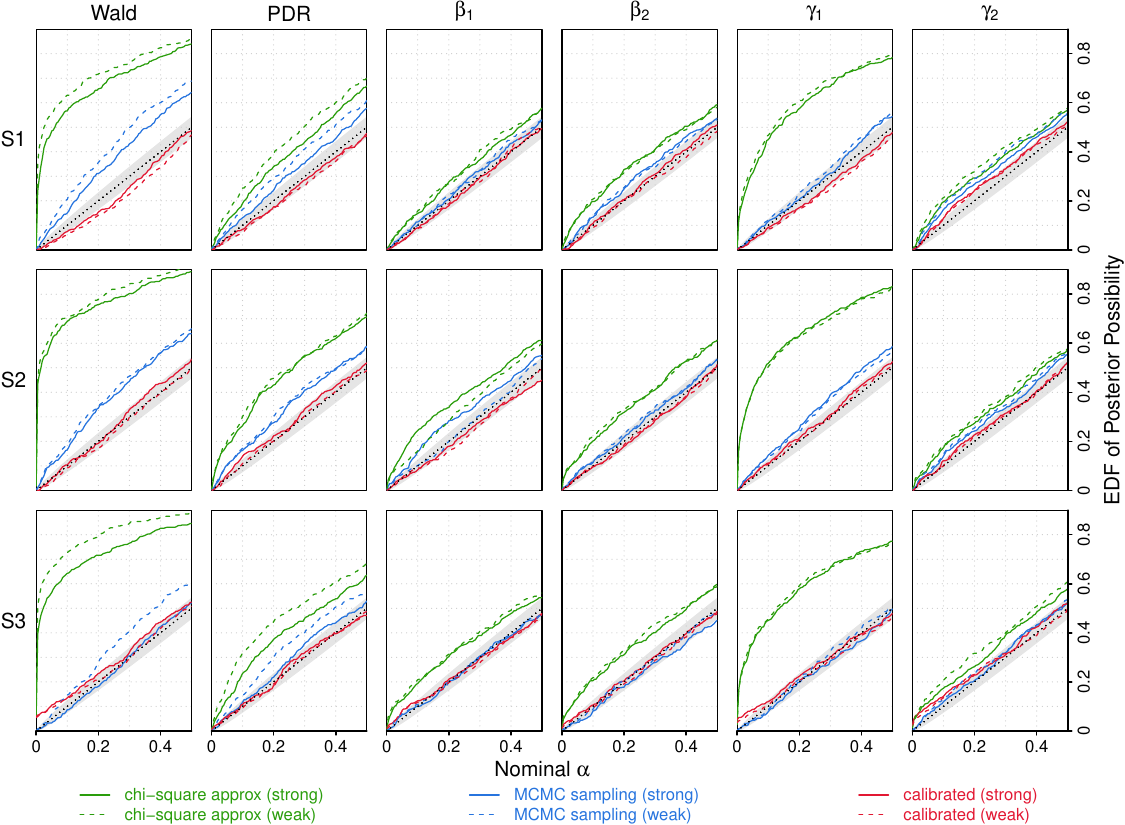}
  \caption{Summary of Study 1: $m = 10$ design variables. Rows of the graphical table represent three parameter-generating scenarios (S1--S3). Columns represent six types of test statistics: the first two columns correspond to the Wald and posterior density ratio (PDR) statistics for simultaneous inference of all parameters, and the remaining four columns correspond to the marginal Wald statistics for selected parameters ($\beta_1$, $\beta_2$, $\gamma_1$, and $\gamma_2$). Six empirical distribution functions (EDFs) of posterior possibilities are presented in each panel. Colors are used to contrast results based on chi-square approximation (green), Markov chain Monte Carlo (MCMC) sampling (blue), and the proposed calibration algorithm (red). Line types are used to distinguish strong ($t_5(0, .5^2)$; solid) and weak ($t_5(0, 25^2)$; dashed) priors. The diagonal dotted lines in each panel indicates exact uniformity; a 95\% normal-approximation, pointwise Monte Carlo confidence band is shown by the gray area. EDFs above the diagonal signifies liberal and thus invalid inference, while EDFs below the diagonal implies conservative and thus valid inference.}
  \label{fig:summ2}
\end{figure}

Results for $m = 3$ are reported in Figure \ref{fig:summ1}. Even though neither prior was correctly specified in Scenario 1, uncalibrated Bayesian inference via MCMC sampling exhibits acceptable performance across the six test statistics, showing only slightly liberal results when the overall Wald statistic and the marginal Wald statistic for $\beta_2$ are in use. Meanwhile, chi-square approximations were noticeably liberal across most test statistics, except when marginal inference is made for the slope parameter $\gamma_2$ of the log-scale function. Performance of uncalibrated Bayesian inference deterioriates in Scenario 2 with fixed true parameters. Both chi-square and MCMC approximations may yield liberal inference, particularly with the overall Wald statistic and the marginal Wald statistic for $\beta_1$. Furthermore, the weakly informative prior (with a scale of 25) slightly outperforms the strongly informative prior (with a scale of .5) in terms of validity. Under Scenario 3, where the strongly informative prior was correctly specified, all candidate methods performed adequately, except for the chi-square approximation applied to the overall Wald statistics. In contrast to the mixed performance of chi-square approximation and MCMC sampling, the calibrated posterior inference consistently maintains validity across almost all parameter-generating scenarios and types of statistics.

Discrepancies among the candidate methods become more salient when the number of design variables is large ($m = 10$; see Figure \ref{fig:summ2}). In Scenario 1, inference based on asymptotic chi-square approximations is consistently liberal; this is particularly severe for simultaneous inference and marginal inference for $\gamma_2$. MCMC sampling generally reduces this liberalism but remains problematic for simultaneous inference and the marginal inference of $\beta_2$ and $\gamma_2$. Moreover, while the two priors yield comparable results for marginal inference, the weakly informative prior is notably more liberal for simultaneous inference. Under Scenario 2 with fixed true parameters, uncalibrated Bayesian inference demonstrates a pattern similar to, but slightly more liberal than, that of Scenario 1. In line with Bayesian validity (\ref{eq:bcalib}), the correctly specified strong prior yields valid inference under Scenario 3 when posterior possibilities are approximated by MCMC sampling. While the weakly informative prior largely preserves validity for marginal inference, the corresponding simultaneous inference is considerably liberal. Meanwhile, the asymptotic chi-square approximation is unacceptably poor under Scenario 3. Similar to the $m = 3$ conditions, calibrated inference remains largely valid across most, if not all, parameter-generating scenarios and test statistics when $m = 10$.

In addition, we performed further simulations for the Gaussian location-scale regression, focusing on the impact of model misspecification and the effect of calibration on statistical power when testing a null effect. Detailed findings are available in the Supplementary Materials.

\color{blue1}
\subsection{Study 2: Gaussian Linear Factor Analysis}

\subsubsection{Data Generation}

Covariance structure models have been widely used for interpreting correlational patterns among observed response variables \citep{Bollen1989, Joreskog1970}. Let $\ZZ = (\ZZ_1, \dots, \ZZ_n)\t\in\real^{n\times m}$ be independent and identically distributed (i.i.d.) sample data, in which each $m$-dimensional response vector $\ZZ_i\sim \sN\left(\mmu, \SSigma(\ttheta)\right)$ with a mean vector $\mmu\in\real^m$ and a covariance matrix $\SSigma(\ttheta)\in\real^{m\times m}_+$ parameterized by $\ttheta\in\real^q$. In our numerical illustration, we consider a one-dimensional common factor model \citep{Joreskog1969}:
\begin{equation}
  \ZZ_i = \mmu + \llambda\eta_i + \eepsilon_i,\ \eta_i\sim\sN(0, \psi),\ \eepsilon_i\sim\sN\left({\bf 0}, \diag(\uupsilon)\right).
  \label{eq:dge3fac}
\end{equation}
In (\ref{eq:dge3fac}), $\mmu\in\real^m$ is the intercept vector, $\llambda = (\lambda_1, \dots, \lambda_m)\t\in\real^m$ is the factor loading vector, $\psi\ge 0$ denotes the common-factor variance, and $\uupsilon = (\upsilon_1, \dots, \upsilon_m)\t\in\real^m_+$ collects unique factor variance parameters. To identify the model, we require the common factor $\eta_i\in\real$ to be independent of the unique factors $\eepsilon_i\in\real^m$, the common factor to be standardized ($\psi = 1$), and the first factor loading to be strictly positive ($\lambda_1 > 0$). The following covariance matrix for the response variables are implied by the common factor model:
\begin{equation}
  \SSigma(\ttheta) = \llambda\llambda\t + \diag(\uupsilon).
  \label{eq:cov3fac}
\end{equation}
To avoid bounding the parameter space, we log-transformed all the positive parameters: Let $\zeta = \log\lambda_1$ be the logarithm of the first factor loading, and $\omega_j = \log\upsilon_j / 2$, $j = 1,\dots, m$, be the log SD of the $j$th unique factor variance. We collect these unbounded parameters in the parameter vector $\ttheta = (\zeta, \lambda_2, \dots, \lambda_m, \omega_1, \dots, \omega_m)\t$, whose dimension is $q = 2m$. I.i.d. normality of $\ZZ_i$, $i = 1,\dots, n$, implies that the sample cross-product matrix, $\YY = \sum_{i=1}^n(\ZZ_i - \bar\ZZ)(\ZZ_i - \bar\ZZ)\t$, where $\bar\ZZ$ denotes the sample mean vector, follows $\mathsf{Wish}(\SSigma(\ttheta), n - 1)$, a Wishart distribution with a scale matrix $\SSigma(\ttheta)$ and degrees of freedom $n - 1$. 
Because the cross-product matrix $\YY$ is a sufficient statistic for the covariance structure $\SSigma(\ttheta)$, we treat $\YY$ as data throughout the simulation study. A direct data generating algorithm for $\YY$ is $\YY = \gg(\UU, \ttheta) = \SSigma(\ttheta)^{1/2}\UU\SSigma(\ttheta)^{1/2}$, where $\UU\sim\mathsf{Wish}(\II_{m\times m}, n - 1)$.

The two design factors of the simulation study are the number of response variables ($m = 5$ and 15) and the parameter generating scenarios (Scene 1--3). The sample size is fixed at $n = 100$. In Scene 1, the communalities of response variables were randomly sampled from $\mathsf{Unif}[.2, .8]$ across replications, spanning low to high communalities \citep{MacCallumEtAl1999}. While the common factor variance was set to one, the unique variances were determined such that all response variables have unit variances. The induced distribution for $\TTheta$ serves as $\pr_{\TTheta}^*$. In Scene 2, all response variables have fixed low communalities (.3). In Scene 3, all response variables have fixed high communalities (.7). Common and unique factor variances are determined similar to Scene 1. The parameter generating distribution $\pr_{\TTheta}^*$ in Scenes 2 and 3 are Dirac measures (i.e., point mass concentrated at the true parameters). 

\subsubsection{Tuning Details and Evaluation Criteria}

Similar to Study 1, we imposed independent $t$-priors on the factor loadings and unique factor SDs. To ensure the positivity of the first factor loading, $\lambda_1$, and the unique SDs, $\sqrt{\upsilon_j}$, $j = 1,\dots, m$, their priors were truncated at 0, resulting in half-$t$ priors. We fixed the degrees of freedom for the $t$-priors at 5, and set the scale parameters to 1 and 25 to represent strong and weak \textit{a priori} information, respectively.

All computational algorithms were tuned in the same fashion as in Study 1. The only exception is that the expected Hessian of the log-posterior was used in place of the observed Hessian for optimization and defining the Wald statistics. The MATLAB code for this simulation is again available at \href{https://github.com/yliu87/CalibBayes}{https://github.com/yliu87/CalibBayes}.

We contrasted uncalibrated posterior contours, obtained using both chi-square approximation and MCMC sampling, with their calibrated counterparts computed via the proposed algorithm. Results were summarized for the overall Wald and PDR statistics, as well as the marginal Wald statistics for selected parameters: $\zeta$, $\lambda_2$, $\omega_1$, and $\omega_2$. Due to the positivity constraint on the first factor loading, inference for the parameters of the first and second response variables could differ. EDF plots of the contour functions, similar to Figures \ref{fig:summ1} and \ref{fig:summ2}, were generated to evaluate inferential validity.

\subsubsection{Results}

\begin{figure}[!t]
  \centering
  \includegraphics[width=\textwidth]{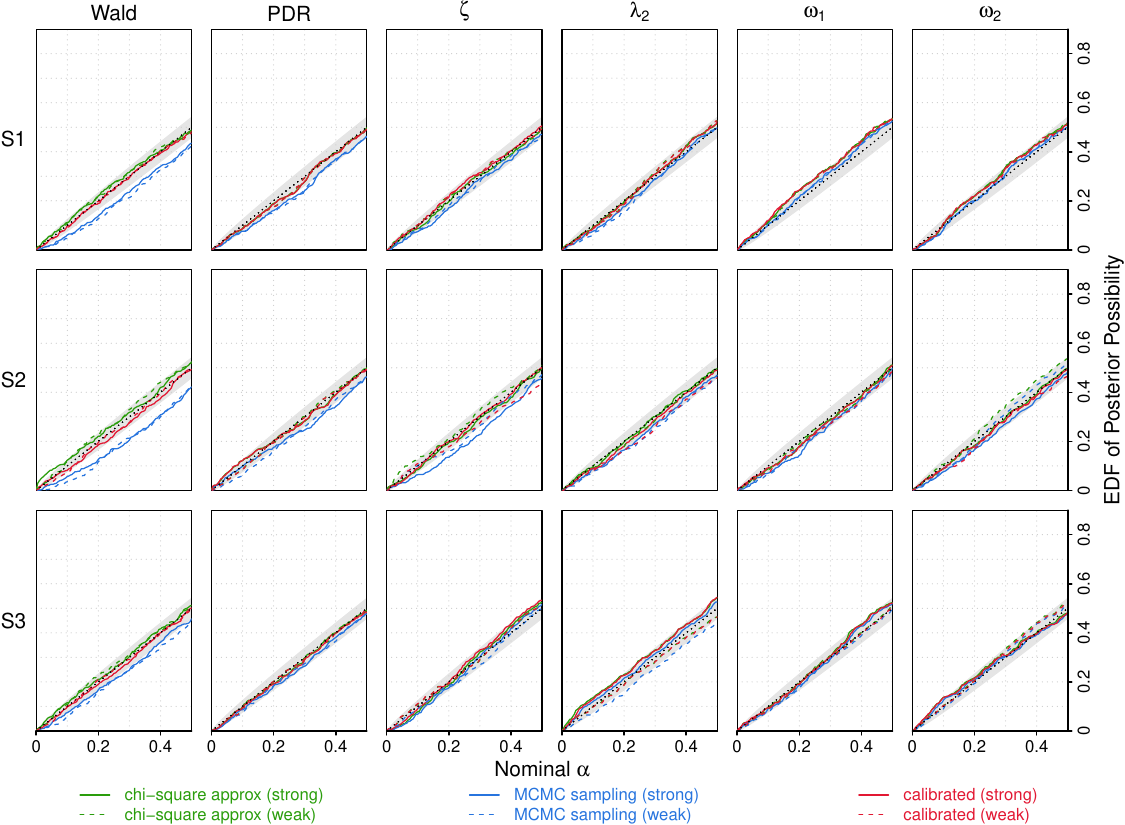}
  \caption{Summary of Study 2: $m = 5$ response variables. Rows of the graphical table represent three parameter-generating scenarios (S1--S3). Columns represent six types of test statistics: the first two columns correspond to the Wald and posterior density ratio (PDR) statistics for simultaneous inference of all parameters, and the remaining four columns correspond to the marginal Wald statistics for selected parameters ($\zeta$, $\lambda_2$, $\omega_1$, and $\omega_2$). Six empirical distribution functions (EDFs) of posterior possibilities are presented in each panel. Colors are used to contrast results based on chi-square approximation (green), Markov chain Monte Carlo (MCMC) sampling (blue), and the proposed calibration algorithm (red). Line types are used to distinguish strong ($t_5(0, 1)$; solid) and weak ($t_5(0, 25^2)$; dashed) priors. The diagonal dotted lines in each panel indicates exact uniformity; a 95\% normal-approximation, pointwise Monte Carlo confidence band is shown by the gray area. EDFs above the diagonal signifies liberal and thus invalid inference, while EDFs below the diagonal implies conservative and thus valid inference.}
  \label{fig:summ3}
\end{figure}

\begin{figure}[!t]
  \centering
  \includegraphics[width=\textwidth]{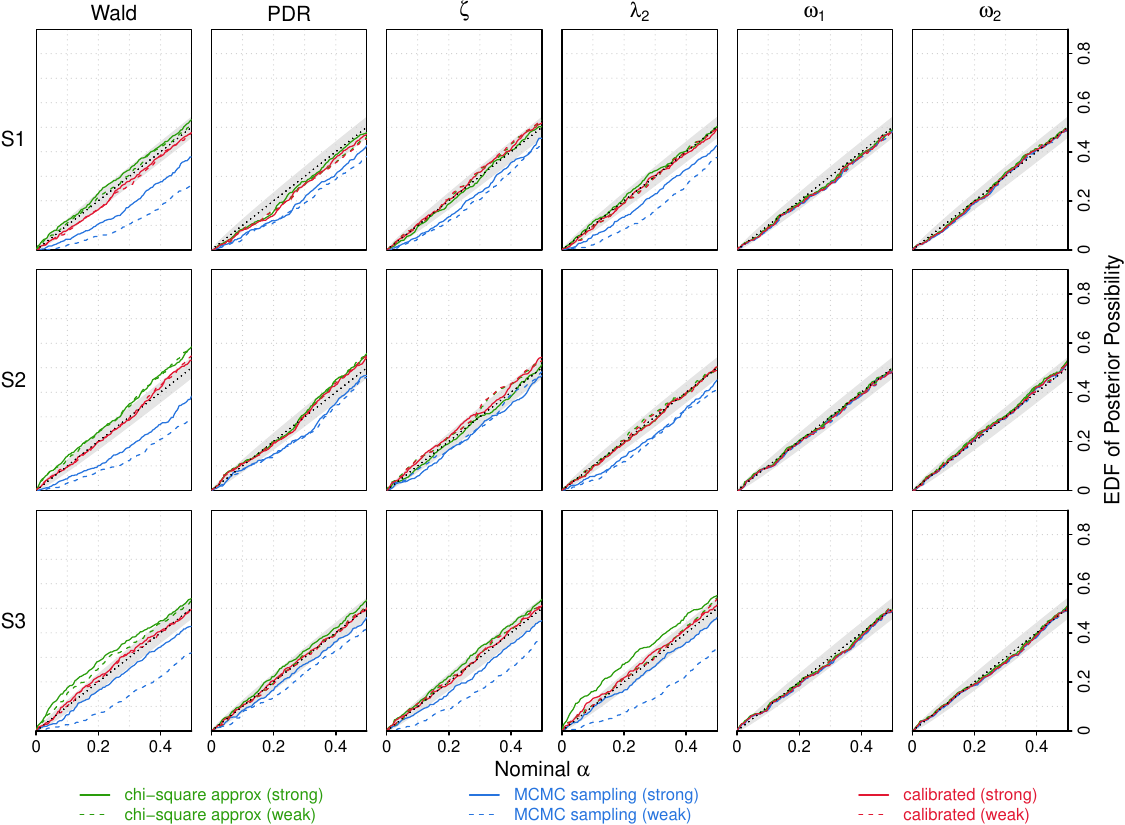}
  \caption{Summary of Study 2: $m = 15$ response variables. Rows of the graphical table represent three parameter-generating scenarios (S1--S3). Columns represent six types of test statistics: the first two columns correspond to the Wald and posterior density ratio (PDR) statistics for simultaneous inference of all parameters, and the remaining four columns correspond to the marginal Wald statistics for selected parameters ($\zeta$, $\lambda_2$, $\omega_1$, and $\omega_2$). Six empirical distribution functions (EDFs) of posterior possibilities are presented in each panel. Colors are used to contrast results based on chi-square approximation (green), Markov chain Monte Carlo (MCMC) sampling (blue), and the proposed calibration algorithm (red). Line types are used to distinguish strong ($t_5(0, 1)$; solid) and weak ($t_5(0, 25^2)$; dashed) priors. The diagonal dotted lines in each panel indicates exact uniformity; a 95\% normal-approximation, pointwise Monte Carlo confidence band is shown by the gray area. EDFs above the diagonal signifies liberal and thus invalid inference, while EDFs below the diagonal implies conservative and thus valid inference.}
  \label{fig:summ4}
\end{figure}

All optimization and sampling procedures converged successfully across all replications and conditions. Unlike Study 1, we observed that the large-sample chi-square approximation is largely on target, with a few exceptions when the prior is strong (i.e., scale = 1) and the number of response variables is large (i.e., $m = 15$). Similarly, Bayesian inference via MCMC sampling tends to remain valid across all conditions. These findings indicate that common practices in Bayesian analysis are not always problematic. However, MCMC-based probability calculations can become noticeably conservative for factor loadings (i.e., $\zeta$ and $\lambda_2$) and simultaneous inference across all parameters, especially when a weakly informative prior (i.e., scale = 25) is specified.

In contrast, our calibration procedure offers a more robust solution. Across all simulation conditions and both prior specifications, the calibrated possibilities of the true parameter values remain approximately uniformly distributed (i.e., the EDF always stays close to the diagonal). In other words, the calibration procedure not only rectifies the undue liberality of asymptotic inference based on chi-square approximation but also improves efficiency when MCMC-based inference proves excessively conservative.
\color{black}

\section{Empirical Example}

\subsection{Data and Model}

In this section, we use a real-data example to illustrate how calibration results can be presented in practice. We analyze the \texttt{Ginzberg} data set from the R package \texttt{carData} \citep{carData}, which contains responses from $n = 82$ psychiatric patients hospitalized for depression. \yl{The outcome of interest is the patients' self-reported scores on the Beck Depression Inventory, while the predictor is ``fatalism'', measuring the belief that events are predetermined and unavoidable. Both variables were adjusted for relevant covariates (labeled as \texttt{adjdep} and \texttt{adjfatal} in the \texttt{Ginzberg} data frame). To capture nonlinearity in the location and log-scale functions, we modeled them as quartic polynomials. An orthogonal polynomial basis matrix was created using the R command \texttt{poly(adjfatal, degree = 4)}, followed by standardization. In total, our model features $m = 5$ design variables and $q = 10$ parameters. We used the weakly informative, independent $t_5(0, 25^2)$ prior for all parameters.}

\begin{figure}[!t]
  \centering
  \includegraphics[width=0.95\textwidth]{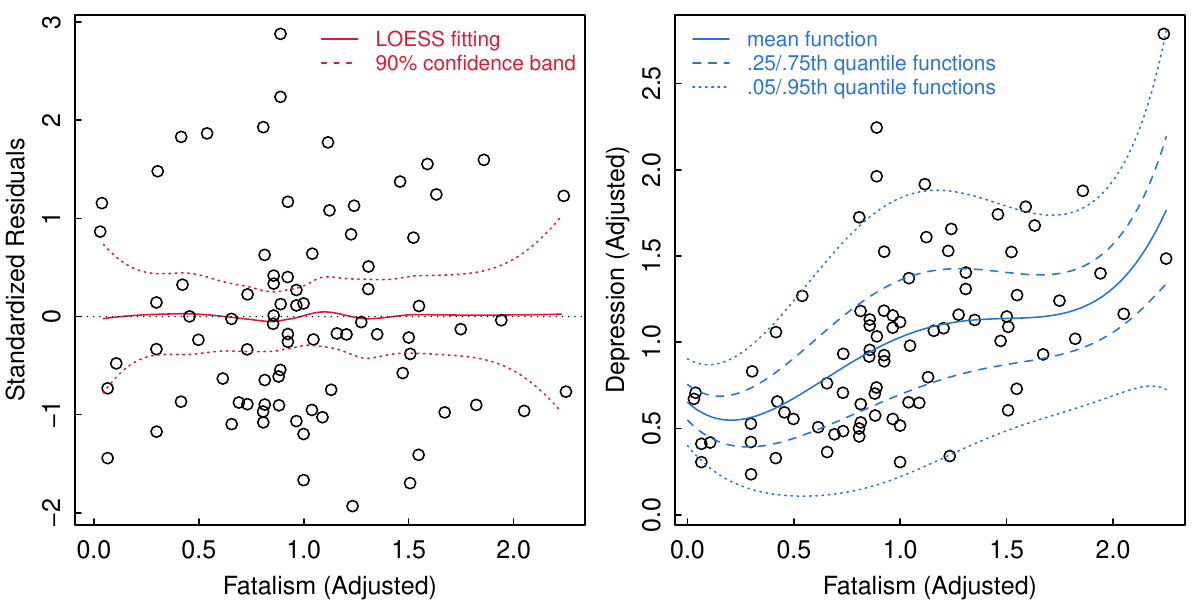}
  \caption{Left: Residual plot. Standardized residuals from the Gaussian location-scale regression are plotted against the predictor fatalism. The solid red curve represents a locally estimated scatterplot smoothing (LOESS) fit, which exhibits no substantial trend. Red dashed curves indicate the corresponding 90\% pointwise confidence band. Right: Fitted mean and quantile functions for depression given fatalism. The solid blue curve superimposed onto the scatterplot depicts the location function. Dashed curves represent the .05th, .25th, .75th, and .95th conditional quantile functions, visualizing the conditional distribution. Under the normality assumption, a conditional quantile is calculated as the conditional mean plus the product of the conditional SD and the corresponding standard normal quantile.}
  \label{fig:exfit}
\end{figure}

\yl{We assessed whether quartic polynomials offer sufficient flexibility to capture the nonlinearity in the location and log scale functions by examining the standardized residuals, $[y_i - \xx_i\t\hat\bbeta(\yy)] / \exp(\xx_i\t\hat\ggamma(\yy))$, $i = 1,\dots, n$, where $\hat\bbeta(\yy)$ and $\hat\ggamma(\yy)$ denote the MAP estimates obtained from the sample data. Plotting these residuals against fatalism (see the left panel of Figure \ref{fig:exfit}) reveals no significant trend, confirming that the regression function is adequately specified. Moreover, the even dispersion of the residuals around zero across the entire domain of fatalism indicates the appropriateness of the quartic log scale function. The right panel of Figure \ref{fig:exfit} displays the fitted mean and quantile functions based on the MAP solution.\footnote{\yl{Under the assumption of normality, any conditional quantile of depression given fatalism is calculated as the conditional mean plus the product of the conditional SD and the corresponding standard normal quantile.}} Although strong nonlinearity is implied by the point (MAP) estimates of the location-scale regression coefficients, we have yet to establish whether this pattern is simply an artifact of sampling variability. In what follows, we compare uncalibrated and calibrated Bayesian inference to determine if they lead to qualitatively different conclusions regarding the shape of the location and log scale functions.}

\subsection{Computational Details}

\yl{When a single data set is analyzed, it is convenient to contrast different inferential procedures by plotting their possibility contours against a pre-specified grid of test statistic thresholds or parameter values. For simultaneous inference based on the Wald and PDR statistics, we chose a grid of 99 threshold levels $\xi_1, \dots, \xi_{99}$, corresponding to the $0.01, 0.02, \dots, 0.99$th quantiles of the $\chi^2_{10}$ distribution. For marginal inference, it is of practical interest to test the significance of trends at various polynomial degrees (i.e., testing $\beta_2,\dots,\beta_5$ for the location regression and $\gamma_2, \dots, \gamma_5$ for the log scale regression). To facilitate testing each focal parameter $\varphi$ against 0, we specified an equally spaced grid $\phi_1, \dots, \phi_{99}$ ranging from $\min\{2\hat\varphi(\yy), 0\}$ to $\max\{0, 2\hat\varphi(\yy)\}$. This construction ensures that 0 is always an endpoint of the grid and that the grid is centered at the MAP estimator $\hat\varphi(\yy)$. 

To improve computational efficiency, we adopted a ``warm start'' strategy for calibration. We began with the threshold or parameter value closest to the MAP estimator. In this initial run, 10000 burn-in iterations were executed, followed by 10000 iterations to obtain the averaged solution. We then performed calibration along the grid sequence, moving away from the MAP solution. For each step, we obtained starting values by retracting the final solution from the previous step onto the current manifold and performed 10000 SPRSA iterations. For marginal inference, this process was carried out toward the left and right tails, respectively. Accordingly, the initial learning rate $\underline a$ is reduced to .1 for simultaneous inference and to .001 for marginal inference. Other tuning aspects of the SPRSA algorithm remained the same as in the simulations.} In addition to chi-square approximations and calibration, we also evaluated uncalibrated posterior probabilities using MCMC sampling, tuned exactly as in the earlier MC experiment.

\subsection{Results}

\begin{figure}[!t]
  \centering
  \includegraphics[width=0.95\textwidth]{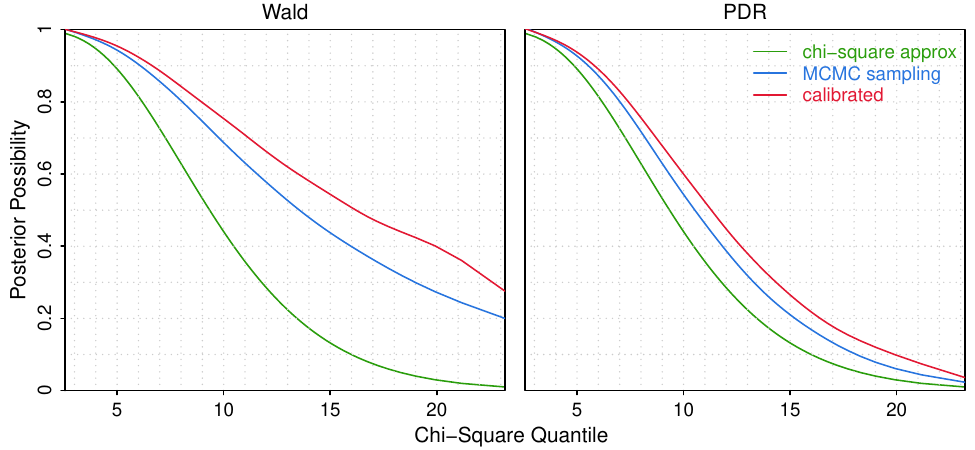}
  \caption{Posterior possibilities for simultaneous inference. Results for the Wald and posterior density ratio (PDR) statistics are displayed in two separate panels. Results based on the chi-square approximation, Markov chain Monte Carlo (MCMC) sampling, and calibration are displayed in green, blue, and red, respectively.}
  \label{fig:exjoint}
\end{figure}

\begin{figure}[!t]
  \centering
  \includegraphics[width=\textwidth]{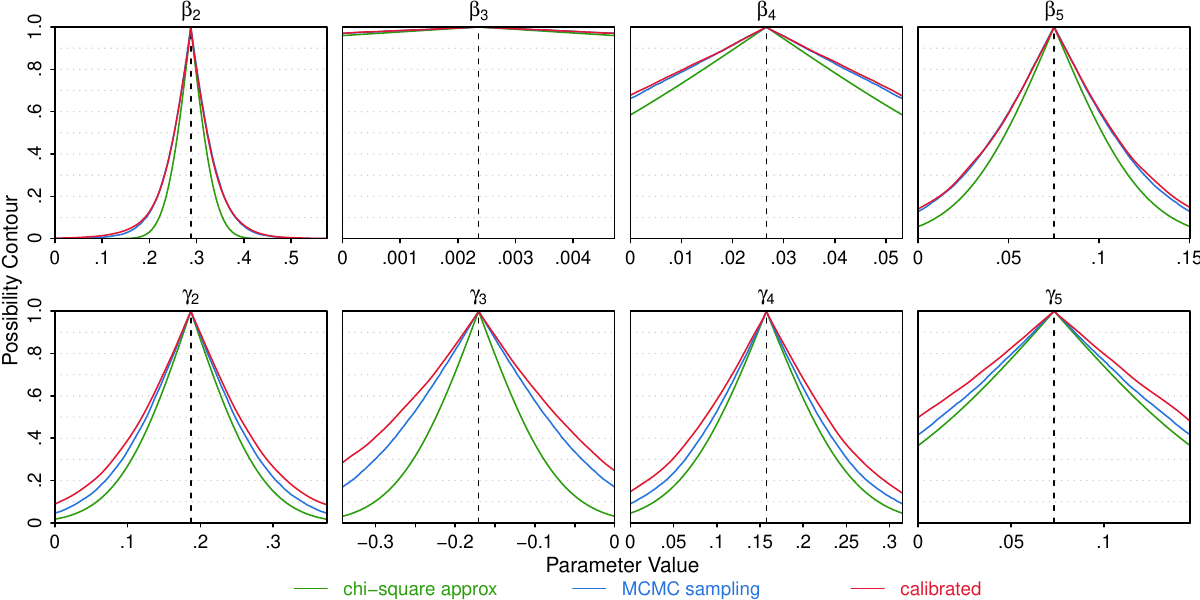}
  \caption{Posterior possibilities for marginal inference. Each panel corresponds to a single focal parameter. Results based on the chi-square approximation, Markov chain Monte Carlo (MCMC) sampling, and calibration are displayed in green, blue, and red, respectively. The vertical dashed line in each panel marks the maximum a posteriori estimate.}
  \label{fig:exmarg}
\end{figure}

The results for the Wald and PDR statistics are presented in the respective panels of Figure \ref{fig:exjoint}. For each statistic, we visualize possibility contours derived from the chi-square approximation (green), MCMC sampling (blue), and SA-based calibration (red) against $\chi^2_{10}$ quantiles.\footnote{\color{blue1}In both Figures \ref{fig:exjoint} and \ref{fig:exmarg}, the possibility contours generated by MCMC sampling and SA-based calibration were smoothed using the R function \texttt{smooth.spline} before plotting.} We observed that MCMC sampling yields considerably more conservative inference than the chi-square approximation. Meanwhile, calibrated inference tends to be even more conservative, dominating the uncalibrated posterior contours across all threshold levels. Such discrepancy is more pronounced for the Wald statistic than for the PDR statistic. These results mirror our findings in the MC experiment, suggesting caution that uncalibrated Bayesian inference may not be valid in small-sample applications of Gaussian location-scale regression.

\yl{Figure \ref{fig:exmarg} presents the marginal posterior possibility contours for the eight polynomial coefficients. The uncalibrated possibility contours derived from the chi-square approximation are generally narrower than those based on MCMC sampling, which are, in turn, notably narrower than the calibrated contours. Consequently, inference is most liberal under the large-sample approximation and most conservative following calibration. Moreover, these three methods yield qualitatively different results when testing polynomial trends, leading to divergent conclusions regarding the shapes of the location and scale functions. To facilitate interpretation, we categorize possibility values (i.e., $p$-values) in the intervals $[0, .05)$, $[.05, .1)$, and $[.1, 1]$ as indicating strong, moderate, and weak evidence, respectively, for a trend of a given polynomial degree. Applying this rule of thumb, the large-sample chi-square approximation indicates strong evidence for a linear trend ($\beta_2$) and moderate evidence for a quartic trend ($\beta_5$) in the location function. It also shows strong evidence for linear, quadratic, and cubic trends ($\gamma_2$, $\gamma_3$, and $\gamma_4$) in the log scale function. In contrast, MCMC-based Bayesian inference offers only weak support for a quartic trend in the location function and marginally moderate support for a cubic trend in the log scale function; however, its support for heterogeneous error variances remains strong due to a significant $\gamma_2$ coefficient at $\alpha = 0.05$. Under calibrated inference, support for a linear trend in the log scale function is further reduced to moderate. In other words, a simple linear regression with homogeneous error variance is largely sufficient to characterize the relationship between fatalism and depression. Overall, these discrepancies highlight that our verdict on nonlinearity can be heavily affected by the choice of inferential procedure, and that uncalibrated Bayesian inference---even with a weak prior---may lead to substantial overfitting.}

 \color{black}

\section{Discussion}

Bayesian statistics is popular among psychologists for its intuitive uncertainty quantification, broad applicability to diverse modeling settings, and sometimes strong performance with small samples \citep[e.g.,][]{Depaoli2021, MuthenAsparouhov2012, VanDeSchootEtAl2021}. However, drawing on the crucial notion of Bayesian validity, we demonstrate that Bayesian methods can be unreliable over repeated samples when the specified prior for inference mismatches the true parameter-generating prior. Since data analysts rarely have information about this true mechanism in practice, we offer a safer alternative: calibrating posterior-based inference to achieve frequentist validity, a stronger requirement that guarantees Bayesian validity with any parameter-generating prior. 
To solve the calibration problem, we develop an SPRSA algorithm that integrates manifold optimization with gradient-free SA. We then report an MC experiment concerning Gaussian location-scale regression. We show that standard Bayesian inference with a popular prior specification can be invalid depending on the credible region type, the number of response variables, and the true parameter-generating mechanism. In contrast, the calibrated Bayesian inference achieves validity in all simulated conditions. Additionally, we demonstrate that the SPRSA algorithm is scalable to realistic problem sizes common in psychological applications. Suggested graphical displays of calibration results were also provided with a real-data analysis.

We highlight two philosophical implications of this work. First, we recognize that Bayesian priors are rarely regarded as literal descriptions of the probabilistic mechanism that selects the truth; rather, they serve as tools to regularize estimation and formally quantify uncertainty. While one can proceed with Bayes' rule using any personal or default prior, the resulting uncertainty quantification can be misleading in the long run. This is demonstrated theoretically through the proof of the FCT \citep{BalchEtAl2019} and empirically in our MC experiment. Consequently, we consider rigorous evaluations of long-run performance imperative. Second, we emphasize that our notions of validity and calibration are fundamentally Fisherian rather than Neyman-Pearson. Specifically, we rely on a conservative decision rule to evaluate hypotheses: we reject a hypothesis only when its possibility is low, and we accept it when the possibility of its complement is low. While establishing validity guarantees Type I error control, it may concurrently inflate the Type II error rate (i.e., reduce statistical power). This trade-off should be carefully weighed during research planning.

The present study has several limitations that should be addressed in future research. First, our simulations were confined to two families of statistical models (i.e., Gaussian location-scale regression and Gaussian linear factor analysis). Given the broad application of Bayesian methods, comprehensive MC experiments are needed to assess how often popular prior configurations lead to invalid inference, thereby highlighting the general necessity for calibration. Second, it is well known that MAP estimation is generally not invariant to reparameterization. Hence, a limitation of the proposed approach is that calibration results may change when imposing priors on transformations of model parameters. Additionally, it remains unclear how to execute calibration in cases of parameter expansion, where priors are specified for an over-parameterized working model. Third, as demonstrated by the additional simulations in the Supplementary Materials, Bayesian inference generally lacks robustness to likelihood misspecification. While goodness-of-fit assessment and model modification are often recommended in practice, partially-specified likelihood offers an alternative solution that can be explored in future work \citep{Martin2022c}. Fourth, the use of Wald and PDR statistics entails repeatedly finding MAP solutions throughout SPRSA iterations, which can be computationally expensive for complex models. Future work may explore alternative FD gradient estimates or test statistics to further alleviate the computational burden. Finally, our method assumes a differentiable $p$-value function, precluding its direct application to discrete data problems. One promising resolution is to add a random perturbation to the test statistic, thereby forcing its distribution to be continuous.

\appendix

\section{Details of the Calibration Algorithm}
\label{a:alg}

Algorithm \ref{alg:sprsa} presents pseudocode of the proposed SPRSA algorithm for solving (\ref{eq:optim}). The algorithm performs gradient ascent iterations on the differentiable submanifold $\partial D_\xi(\yy)$, in which the exact Riemannian gradient of the $p$-value function $\pi_\yy(\ttheta)$ is approximated by a noisy finite-difference (FD) estimate. To improve the rate of convergence, the iterates are averaged by the recursive formula of \citet{TripuraneniEtAl2018}. Further details regarding Riemannian optimization and the convergence of the SPRSA algorithm can be found in the Supplementary Materials.


\subsection{Riemannian Gradient Ascent}

\begin{algorithm}[!b]
  \caption{\texttt{SPRSA}: Simultaneous perturbation Riemannian stochastic approximation}
  \begin{algorithmic}[1]
    \Require Observed data $\bf y\in\cY$, starting values of parameters $\ttheta^{(1)}\in\real^q$, threshold of test statistic $\xi\in\real$, number of iterations $K > 0$, tuning constants $a_1 > 0$, $\beta\in(\delta + 1/2, 1]$, $\gamma > 0$, and $\delta\in(0, 1/2)$
    \State Initialize the average $\bar\ttheta^{(1)} = \ttheta^{(1)}$
  \For{$k = 1, \dots, K$}
  \State Compute learning rate $a_k = \underline ak^{-\underline b}$ and finite-difference rate $c_k = c k^{\underline d}$
  \State Compute noisy Riemannian gradient $\widehat{\grad\pi_{\yy}}(\ttheta^{(k)}) = \mathtt{RiemGradFD}(\ttheta^{(k)}, \yy, c_k)$
  \State Update the parameters $\ttheta^{(k + 1)} = \RR(\ttheta^{(k)}, a_k\widehat{\grad\pi}_{\yy}(\ttheta^{(k)}))$
  \State Update the average $\bar\ttheta^{(k + 1)} = \RR(\bar\ttheta^{(k)}, k^{-1}\RR^{-1}(\bar\ttheta^{(k)}, \ttheta^{(k + 1)}))$\label{ln:avg}
  \EndFor
  \State Return $\bar\ttheta^{(K+1)}$
  \end{algorithmic}
  \label{alg:sprsa}
\end{algorithm}

The Riemannian gradient specifies the local steepest ascent direction and is computed as the orthogonal projection of the ambient gradient to the tangent space of the submanifold. In our problem (\ref{eq:optim}), the submanifold $\partial D_\xi(\yy)$ is implicitly defined as the level set of the test statistic $T(\yy, \ttheta)$; hence, the tangent space at $\ttheta$, denoted $\cT_{\ttheta}\partial D_\xi(\yy)$, is the $(q - 1)$-dimensional null space of $\nabla_{\ttheta}T(\yy, \ttheta)\in\real^q$. The Riemannian gradient is then
\begin{equation}
  \grad\pi_{\yy}(\ttheta) = \proj_{\cT_{\ttheta}\partial D_\xi(\yy)}\nabla_{\ttheta}\pi_\yy(\ttheta) = \underbrace{\left[\II_{q\times q} - \frac{\nabla_{\ttheta} T(\yy, \ttheta)\nabla_{\ttheta} T(\yy, \ttheta)\t}{\nabla_{\ttheta} T(\yy, \ttheta)\t\nabla_{\ttheta} T(\yy, \ttheta)}\right]}_{=:\MM(\ttheta)}\nabla_{\ttheta}\pi_\yy(\ttheta),
  \label{eq:riemgrad}
\end{equation}
in which $\II_{q\times q}$ is a $q$-dimensional identity matrix. The bracketed term on the right-hand side of (\ref{eq:riemgrad}) provides the exact expression of the $q\times q$ matrix $\MM(\ttheta)$ in (\ref{eq:sprsak}).

Let $\RR:\real^q\times \cT_{\ttheta}\partial D_\xi(\yy)\to\partial D_\xi(\yy)$ be a \emph{retraction} at $\ttheta\in\partial D_\xi(\yy)$ that satisfies the centering condition, $\RR(\ttheta, \mathbf{0}_q) = \ttheta$, and the local rigidity condition, $\nabla_t\RR(\ttheta, t\hh)|_{t = 0} = \hh$ \citep[][Definition 4.1.1]{AbsilEtAl2008}. By definition, a retraction maps a tangent vector $\hh\in\cT_{\ttheta}\partial D_\xi(\yy)$ onto the manifold and preserves the direction of $\hh$ in the vicinity of $\ttheta$. A convenient method to define a retraction, known as a \emph{projection-like} retraction, is to find the intersection of the manifold with a one-dimensional linear subspace $\cL(\ttheta, \hh)\subset\real^q$ that passes through the point $\ttheta + \hh$ and is transverse to the tangent space $\cT_{\ttheta}\partial D_\xi(\yy)$ \citep{AbsilMalick2012}. In our problem, we consider the linear space $\cL(\ttheta, \hh) = \{\vvartheta\in\real^q: \hat\ttheta(\yy) + x[\ttheta + \hh - \hat\ttheta(\yy)],\ x\in\real\}$, which is uniquely determined by the MAP estimator $\hat\ttheta(\yy)$ and the current location $\ttheta$. The corresponding retraction is then obtained by solving $x$ from 
\begin{equation}
  \RR(\ttheta, \hh) = \hat\ttheta(\yy) + \chi(\yy, \ttheta)[\ttheta + \hh - \hat\ttheta(\yy)],
  \label{eq:retr}
\end{equation}
in which $\chi(\yy, \ttheta)\in\real$ is the solution of $x$ to $T(\yy, \hat\ttheta(\yy) + x[\ttheta + \hh - \hat\ttheta(\yy)]) = \xi$.

\subsection{Simultaneous Perturbation Gradient Approximation}

\begin{algorithm}[!t]
  \caption{\texttt{RiemGradFD}: Noisy Riemannian gradient by simultaneous perturbation}
  \begin{algorithmic}[1]
    \Require Current iterate $\ttheta\in\real^q$, observed data $\bf y\in\cY$, finite difference step size $c > 0$
  \State Sample from $\pr_\UU$ and denote the realization by $\uu$
  \State Compute noisy ambient gradient $\widehat\pi_\yy(\ttheta; \uu)$  by (\ref{eq:spgrad})
  \State Return noisy Riemannian gradient $\widehat{\grad\pi_{\yy}}(\ttheta) = \proj_{\cT_{\ttheta}\partial D_\xi(\yy)}\widehat\pi_\yy(\ttheta; \uu)$
  \end{algorithmic}
  \label{alg:riemgrad}
\end{algorithm}

Using FD gradient approximation in SA can be traced back to \citet{KieferWolfowitz1952}. In our problem, the objective $\pi_{\yy}(\ttheta)$ is an expectation with an intractable gradient. A noisy FD estimate for the partial derivative of the $r$th parameter, $r = 1,\dots, q$, can be expressed as 
\begin{equation}
  (2c)^{-1}\big[1\{T(\YY, \ttheta + c\ee_r)\ge T(\yy, \ttheta + c\ee_r)\}- 1\{T(\YY, \ttheta - c\ee_r)\ge T(\yy, \ttheta - c\ee_r)\}\big]
  \label{eq:finite}
\end{equation}
for some small perturbation $c > 0$, in which $\ee_r$ is an elementary vector with 1 on the $r$th element and 0 elsewhere. Sending $c\to 0$ at a slow rate along the SA iterations leads to a standard Kiefer-Wolfowitz algorithm. However, the estimate (\ref{eq:finite}) suffers from a ``curse of dimensionality'': more evaluations of the test statistics are required as the number of parameters increases. To address these issues, we apply the technique of simultaneous-perturbation FD originally proposed by \citet{Spall1992}.

  Let $\YY = \gg(\UU, \ttheta)$ be a data-generating algorithm, in which the random components $\UU\sim\pr_{\UU}$ and the distribution $\pr_{\UU}$ is completely known. At iteration $k$, the simultaneous-perturbation FD estimator for the ambient gradient $\nabla_{\ttheta}\pi_{\yy}(\ttheta^{(k)})$ is defined as
        \begin{equation}
          \begin{aligned}
            \widehat{\nabla_{\ttheta}\pi_{\yy}}(\ttheta^{(k)}) = &\ (2c_k\DDelta_k)^{-1}\big[ 1\{T(\gg(\UU^{(k)}, \ttheta^{(k)} + c_k\DDelta_k), \ttheta^{(k)} + c_k\DDelta_k)\ge T(\yy, \ttheta^{(k)} + c_k\DDelta_k)\}\\  
            &\ - 1\{T(\gg(\UU^{(k)}, \ttheta^{(k)} - c_k\DDelta_k), \ttheta^{(k)} - c_k\DDelta_k)\ge T(\yy, \ttheta^{(k)} - c_k\DDelta_k)\}\big],
          \end{aligned}
          \label{eq:spgrad}
        \end{equation}
        in which $\UU^{(k)}\sim\pr_{\UU}$, and $\DDelta_k\in\real^{q}$ is a vector of $q$ independent Rademacher random variables (i.e., taking values $-1$ or 1 with 50/50 chance). Lemma 1 of \citet{Spall1992} establishes that, when $c_k\to 0$ as $k\to\infty$, the bias of (\ref{eq:spgrad}) in estimating $\nabla_{\ttheta}\pi_{\yy}(\ttheta^{(k)})$ is of order $o(c_k^2)$. Compared to (\ref{eq:finite}), which demands two evaluations of the test statistic $T$ for every coordinate of the parameter vector, the simultaneous-perturbation estimator (\ref{eq:spgrad}) only requires two statistics evaluations in total. The noisy ambient gradient (\ref{eq:spgrad}) is then projected onto the tangent space of the submanifold as a noisy Riemannian gradient. For ease of reference, the simultaneous-perturbation approximation of the Riemannian gradient is summarized in Algorithm \ref{alg:riemgrad}.

\subsection{Averaging}
As detailed in the Supplementary Materials, the SPRSA algorithm converges to a local solution of (\ref{eq:optim}) on the manifold $\partial D_\xi(\yy)$, provided the learning rate sequence $\{a_k\}$ and the FD sequence $\{c_k\}$ satisfy 
\begin{equation}
  \sum_{k = 1}^\infty a_k = \infty,\ \sum_{k=1}^\infty\frac{a_k^2}{c_k^2} < \infty.
  \label{eq:seq}
\end{equation}
Generalizing the Polyak-Ruppert averaging in Euclidean SA \citep{PolyakJuditsky1992, Ruppert1988}, \citet{TripuraneniEtAl2018} showed that averaging a converging sequence of Riemanniann gradient iterates can speed up convergence. A recursive formula to perform online averaging (see also Line \ref{ln:avg} in Algorithm \ref{alg:sprsa}) is
\begin{equation}
  \bar\ttheta^{(k + 1)} = \RR\left(\bar\ttheta^{(k + 1)}, k^{-1}\RR^{-1}(\bar\ttheta^{(k)}, \ttheta^{(k + 1)})\right),
  \label{eq:avgest}
\end{equation}
in which the current averaging estimate is denoted by $\bar\ttheta^{(k)}$, and $\RR^{-1}:\real^q\times\real^q\to\cT_{\ttheta}\partial D_\xi(\yy)$ denotes the inverse retraction operator:
\begin{equation}
  \RR^{-1}(\ttheta, \ttheta') = \frac{[\ttheta' - \hat\ttheta(\yy)]\t\nabla_{\ttheta} T(\yy, \ttheta)}{[\ttheta - \hat\ttheta(\yy)]\t\nabla_{\ttheta} T(\yy, \ttheta)} \left[\ttheta - \hat\ttheta(\yy)\right] - \ttheta' + \hat\ttheta(\yy).
  \label{eq:invretr}
\end{equation}

Achieving practical efficiency with the SPRSA algorithm requires the careful, case-by-case tuning of several aspects: the learning rate sequence $\{a_k\}$, the FD rate sequence $\{c_k\}$, and the total number of iterations $K$. Tuning details of the SPRSA algorithm in our numerical study are provided in the ``Sampling and Tuning Details'' part of the ``Monte Carlo Experiment'' section.
\color{black}

\bibliographystyle{apacite}
\bibliography{CalibBayes}

\clearpage
\includepdf[pages=-]{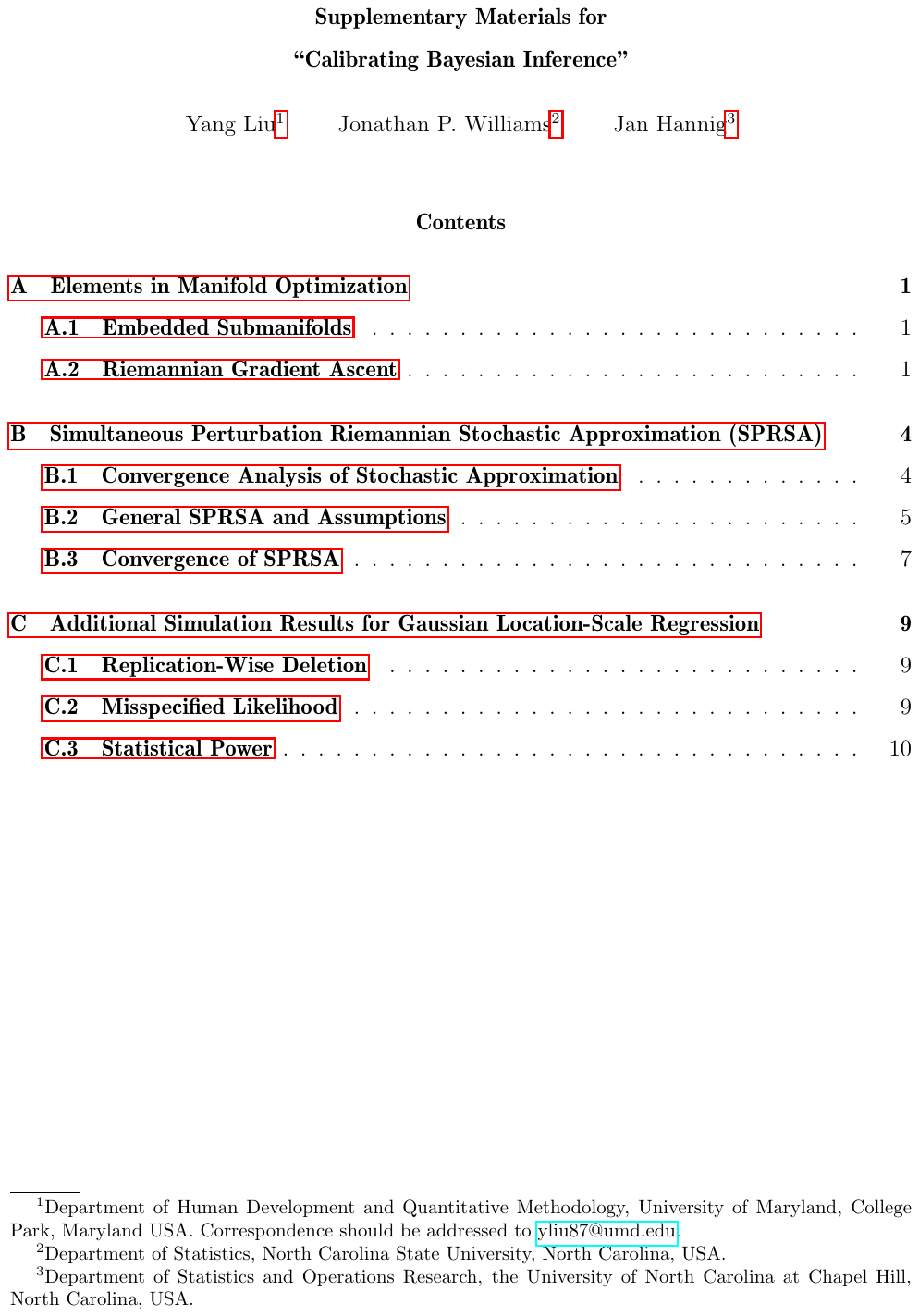}

\end{document}